\documentclass[
 reprint,
 amsmath,
 amssymb,
 aps,nofootinbib,
 superscriptaddress,
 longbibliography,
 balancelastpage
]{revtex4-2}
\usepackage{CJKutf8}

\usepackage[pdftex]{graphicx}
\usepackage{wasysym}
\usepackage{amsfonts}
\usepackage{ifsym}
\usepackage{todonotes}
\usepackage{color}
\usepackage{graphicx}
\usepackage{dcolumn}
\usepackage{bm}
\usepackage{hyperref}
\usepackage{mathtools}
\usepackage{booktabs}
\usepackage[normalem]{ulem}
\hypersetup{
    colorlinks=true,       
    linkcolor=blue,          
    citecolor=magenta,        
    filecolor=red,      
    urlcolor=cyan           
}

\newcommand{\nn}{\nonumber}

\newcommand{\COMMENT}[1]{}

\newcommand{\neqa}{\nonumber\end{eqnarray}}
\newcommand{\la}[1]{\label{#1}}

\renewcommand{\d}{\partial}

\newcommand{\<}{{\langle}}
\renewcommand{\>}{{\rangle}}

\renewcommand{\v}{{\rm v}}

\def\One{1\hskip-.16cm1}

\newcommand{\re}{\relax{\rm I\kern-.18em R}}

\def\su2{{SU(2)}}
\def\eps{{\epsilon}}

\def\[{\left[}
\def\]{\right]}

\def\DD{\mathbb{D}}

\def\({\left(}
\def\){\right)}
\def\[{\left[}
\def\]{\right]}

\def\<{\langle}
\def\>{\rangle}

\def\i2{\frac{i}{2}}

\def\cF{{\cal F}}
\def\cW{{\cal W}}

\def\cC{{\cal C}}

\def\cO{{\cal O}}

\def\cN{{\cal N}}

\def\cP{{\cal P}}

\def\2F1{\,_2{\rm F}_1}

\newcommand{\rrangle}{\rangle \hspace{-.15em} \rangle}
\newcommand{\llangle}{\langle\hspace{-.15em}\langle}

\newcommand{\ii}{\mathrm{i}}
\newcommand{\dd}{\mathrm{d}}

\newcommand{\beq}{\begin{equation}}
\newcommand{\eeq}{\end{equation}}
\newcommand{\beqq}{\begin{equation*}}
\newcommand{\eeqq}{\end{equation*}}
\newcommand\beqa{\begin{eqnarray}}
\newcommand\eeqa{\end{eqnarray}}
\newcommand\beqaa{\begin{eqnarray*}}
\newcommand\eeqaa{\end{eqnarray*}}
\newcommand\bea{\begin{array}}
\newcommand\eea{\end{array}}

\begin{document}
\begin{CJK*}{UTF8}{gbsn}

\title{Universal Constraints for Conformal Line Defects}

\author{Barak Gabai}
\affiliation{Laboratory for Theoretical Fundamental Physics, EPFL, Rte de la Sorge, CH-1015, Lausanne}
\author{Amit Sever}
\affiliation{School of Physics and Astronomy, Tel Aviv University, Ramat Aviv 69978, Israel}
\author{De-liang Zhong (钟德亮)}
\affiliation{Theoretical Physics Group, The Blackett Laboratory, Imperial College London, Prince Consort Road London, SW7 2AZ, UK}

\begin{abstract}

We present a novel framework for deriving integral constraints for correlators on conformal line defects. These constraints emerge from the non-linearly realized ambient-space conformal symmetry. 
To validate our approach, we examine several examples and compare them against existing data for the four-point function of the displacement operator. Additionally, we provide a few new predictions that extend the current understanding of these correlators.

\end{abstract}
\maketitle
\end{CJK*}

\section{Introduction}

The conformal bootstrap is a powerful technique for carving out the space of conformal field theories (CFTs). It is based on the conformal and crossing symmetries of CFT correlation functions; see \cite{Rychkov:2016iqz, Simmons-Duffin:2016gjk, Poland:2018epd, Chester:2019wfx, Hartman:2022zik, Bissi:2022mrs, Rychkov:2023wsd} and references therein. 
This approach extends straightforwardly to conformal defects \cite{Billo:2016cpy,Gadde:2016fbj}. In this paper, we focus on one-dimensional line defects. 
We argue that the correlation functions on these extended operators are subject to a new infinite set of bootstrap constraints, which go beyond those that follow from the conformal and crossing symmetries on a straight line.

The origin of these constraints is simple. Conformal line operators do not exhibit a conformal anomaly, allowing them to be placed along arbitrary smooth paths without breaking conformal symmetry.\footnote{In other words, their expectation value and correlation functions are conformal invariant functionals of their shape.} When expanded around a straight line, the conformal symmetry of the deformed line translates into a set of constraints for integrated correlators. An analogy one can draw is between these constraints and identities that follow from Goldstone's or soft theorems. 

More concretely, a straight conformal line operator is invariant under the $SL(2,{\mathbb R})\times SO(d-1)$ subgroup of the ambient conformal symmetry. Operators localized on the line are classified by this symmetry, which also constrains their correlators. This classification extends to smooth lines, as they are locally straight. Small deformations of the straight line can be parameterized in terms of correlators of operators living on the line. At linear order in the deformation, these considerations imply the existence of the so-called {\it displacement operator}, ${\mathbb D}$, a primary under $SL(2,{\mathbb R})$ with dimension two and transverse spin one. 
At higher orders in the deformation, the deformed line can be expressed in terms of integrated correlators of the displacement operators. Contributions from non-protected operators are excluded due to their anomalous dimensions, making their appearance incompatible with conformal symmetry. Applying a conformal transformation to a deformed line and expanding around the straight line yields an infinite tower of constraints on these integrated correlators.

It is interesting to explore how restrictive these constraints are in determining the expectation value of the line defect. In this paper, we take a first step in this direction by spelling some of these explicitly and checking them against known data at weak and strong coupling. 

The paper is organized as follows. In Section \ref{setup}, we set our conventions for line correlators. Section \ref{constraints} summarizes the constraints on four-point correlation functions on the line. A detailed derivation of these constraints is presented in appendix \ref{derivation app}. In Section \ref{checks}, we provide a non-trivial check that the constraints satisfy, along with a summary of several additional checks we have performed. Future directions are discussed in Section \ref{sec:discussion}. Five appendices provide supplementary material to the main text.

\section{Setup}\la{setup}

We consider a smooth deformation of the straight line 
\beq\la{deformation}
x_\text{straight}(\tau) \quad \mapsto \quad x(\tau)=x_\text{straight}(\tau)+\v(\tau)\,,
\eeq
where $x(\tau)$ is some parametrization of the line and $\v(\tau)$ is the deformation vector, transverse to the line. At leading order in $\v$, the change in the line operator $\cW$ is given by
\beq\la{displacement}
\delta\cW=\int\dd\tau\,|\dot x_\tau|\v_\tau^i\, \cP[{\mathbb D}_i(x_\tau)\cW]\,,
\eeq
where $\dot x_\tau=\d_\tau x(\tau)$ and $i=1,\ldots,d-1$ is a transverse index. Here, $\mathbb{D}_i$ is the displacement operator, whose two-point function is fixed by the conformal symmetry up to an overall constant, $\Lambda$, as
\beq\la{Dtpf}
\llangle \mathbb{D}_i(x) \mathbb{D}_j(0) \rrangle =\frac{\Lambda\,\delta_{ij}}{|x|^4}\,.
\eeq
Here, $\llangle\ldots\rrangle$ denotes correlation functions of operators inserted along the infinite straight line, normalized such that $\llangle \openone \rrangle=1$. 

At higher order in the deformation $\v$, multiple displacement operators can be integrated along the line. The correlation of any odd number of these spinning operators is zero. Hence, the next non-zero correlator is the four-point function
\beq\la{eqn-4ptDisGenForm}
\llangle \mathbb{D}_i(x_1) \mathbb{D}_j(x_2) \mathbb{D}_k(x_3) \mathbb{D}_l(x_4) \rrangle =\frac{\widehat{\mathcal{F}}_{ijkl}(t)}{|x_{14}x_{23}|^{4}}\,,
\eeq
where $x_{ij}=x_i-x_j$ and $x_1<x_2<x_3<x_4$ are four ordered points on the line. Here, 
\beq\la{ccr}
t=\frac{x_{12} x_{34}}{x_{14} x_{23}}\ge0\,,
\eeq
is the $SL(2,\mathbb{R})$ invariant cross-ratio and $\widetilde{\mathcal{F}}_{ijkl}$ can be written as a sum of tensor structures. 

The correlator (\ref{eqn-4ptDisGenForm}) is invariant under cyclic permutation of the four points, which translates into the crossing relation
\beq \label{eqn-CrossingF}
\widehat{\mathcal{F}}_{ijkl}(t) =  t^{-4}\,\widehat{\mathcal{F}}_{jkli}(1/t)\,.
\eeq
We can equally start with the two-point function of a defect operator $\mathbb{O}$ of $SL(2,{\mathbb R})$ conformal dimension $\Delta_{\mathbb{O}}$ on the straight line and then smoothly deform it. For such an operator, the ratio
\beq\la{eqn-4ptDisDisOO}
{\llangle\mathbb{O}_I(x_1)\mathbb{D}_i(x_2) \mathbb{D}_j(x_3)\mathbb{O}_J(x_4) \rrangle\over\llangle \mathbb{D}_1(x_2) \mathbb{D}_1(x_3)\rrangle \llangle\mathbb{O}_1(x_1)\mathbb{O}_1(x_4)\rrangle}
=\widehat{\mathcal{F}}^{\,\mathbb{O} \DD \DD \mathbb{O}}_{IijJ}(t)\,,
\eeq
only depends on the conformal cross ratio $t$ in (\ref{ccr}). If $\mathbb{O}$ transforms under internal or space-time symmetries, then $I$ and $J$ represent the corresponding indices.

The constraints that we consider are most naturally written in terms of the {\it subtracted four-point function}, 
obtained from $\widehat{\mathcal{F}}$ by subtracting the generalized free field contribution $\mathcal{F} \equiv \widehat{\mathcal{F}}- \mathcal{F}_\text{GFF}$, where
\begin{align}
\label{eq:GFFpiece}
   \(\mathcal{F}_\text{GFF}\)^{\mathbb{O} \DD \DD \mathbb{O}}_{IijJ}&=\delta_{ij} \delta_{IJ}\, ,  \\  \(\mathcal{F}_\text{GFF}\)_{ijkl}&=\delta_{il}\delta_{jk} +\delta_{ij}\delta_{kl}\,t^{-4} +\delta_{ik}\delta_{jl}\left(1+t\right)^{-4}\, .\nn
\end{align}

\section{Constraints for 4-point functions}\la{constraints}

A four-point function first appears at the fourth-order variation. Instead, we find it simpler to start with a two-point function of a $SL(2,\mathbb{R})$ primary operator $\mathbb{O}_I$ on the line and apply a second-order variation to it. If $\mathbb{O}={\mathbb D}$, the resulting four-point function corresponds to that of four displacement operators, as given in \eqref{eqn-4ptDisGenForm}.
Otherwise, it takes the form (\ref{eqn-4ptDisDisOO}).

For a generic defect, the dimensions of all defect operators are unrelated. There are, however, examples where this is not the case. For our constraints to hold we have to make the following two assumptions,
\begin{itemize} \label{assume}
\item The is no scalar primary defect operator that is a singlet under internal symmetries and has dimension $3$.
\item There is no non-trivial primary defect operator of dimension $\Delta_{\mathbb O}+1-m$, with non-negative integer $m$, the same charges as $\mathbb O$ and a transverse spin that differs by one unit.
\end{itemize}

We start with the two-point function $\llangle\mathbb{O}_I\mathbb{O}_J\rrangle$ and apply an arbitrary smooth variation to it. We then apply to it a conformal transformation. Under this transformation, the conformal defect transforms into the same defect along the new deformed path. We then expand the result on a straight line. At second order in the deformation and under our first assumption, this leads to two integrations of the form (\ref{displacement}) only. The defect operator $\mathbb O$, on the other hand, transforms. 
At second order in the deformation and under our second assumption, this results in a conformal factor and a spin factor, accounting for the dilatation and transverse rotation of the defect operator respectively. These transformation properties are derived in Section 2 of \cite{Gabai:2023lax} and are adapted to our setup in Appendix \ref{derivation app}.

By applying a Mellin transform to the two integrations,  
one of them can be performed straightforwardly. The remaining integral still depends on an arbitrary deformation vector. Using the fact that an infinitesimal conformal deformation of the line corresponds to a quadratic polynomial, only a limited number of moments of this vector contribute. In this way, we arrive at two types of integrated constraints. 

The first type is homogeneous and takes the form\footnote{These constraints can also be obtained by studying the three-point function $\llangle \mathbb{O} \DD \mathbb{O} \rrangle$ to linear order in the deformation vector $\v$.}
\begin{equation}\label{eqn-Constraint-HomoFinal}
\begin{split}
\int\limits_0^\infty\dd t\[t^2(\mathcal{F}^{\mathbb{O} \DD \DD \mathbb{O}}_{JijI}+\mathcal{F}^{\DD \mathbb{O} \DD \mathbb{O}}_{iIjJ})+(1+t)^2\mathcal{F}^{\mathbb{O} \DD \DD \mathbb{O}}_{JjiI}\] &= 0\, , \\
\int\limits_0^\infty\dd t\,t\[(1+t)\(\mathcal{F}^{\mathbb{O} \DD \DD \mathbb{O}}_{JijI}+\mathcal{F}^{\mathbb{O} \DD \DD \mathbb{O}}_{JjiI}\)-\mathcal{F}^{\DD \mathbb{O} \DD \mathbb{O}}_{iIjJ} \] &= 0\, ,  \\
\int\limits_0^\infty\dd t\[(1+t)^2\mathcal{F}^{\mathbb{O} \DD \DD \mathbb{O}}_{JijI}+t^2\mathcal{F}^{\mathbb{O} \DD \DD \mathbb{O}}_{JjiI}+\mathcal{F}^{\DD \mathbb{O} \DD \mathbb{O}}_{iIjJ}\] &= 0\,,
	\end{split}
\end{equation}
where as before $\mathcal{F} = \widehat{\mathcal{F}}- \mathcal{F}_\text{GFF}$ 
and \footnote{The prefactors in \eqref{funnyFs} are introduced so that when $\mathbb{O}=\DD$ all the $\cF$'s are equal and the constraints can be further simplified, see Appendix \ref{apd:KernelDDDDSimplified}.}
\beq \label{funnyFs}
\widehat{\mathcal{F}}^{\,\DD \mathbb{O} \DD \mathbb{O}}_{iIjJ}(t)\equiv{1 \over (1+t)^4}{\llangle\DD_i(x_1)\mathbb{O}_I(x_2) \mathbb{D}_j(x_3)\mathbb{O}_J(x_4) \rrangle\over\llangle \mathbb{D}_1(x_1) \mathbb{D}_1(x_3)\rrangle \llangle\mathbb{O}_1(x_2)\mathbb{O}_1(x_4)\rrangle}\,.
\eeq

These types of constraints also apply in the case where ${\mathbb O}_I$ and ${\mathbb O}_J$ are replaced by any two defect operators that are not related by symmetry. 
The second type of constraint is inhomogeneous and takes the form 
\footnote{When operators of noninteger dimensions lower than $4$ appear in one of the OPE channels, fractional power divergences should be subtracted.} 
\begin{widetext}
\beq\la{eqn-Constraint-InHomoFinal-1}
\begin{split}
\Lambda\int\limits_{0}^{\infty} \dd t \Bigg[\Big(t^2\big[\mathcal{F}^{\mathbb{O} \DD \DD \mathbb{O}}_{IjiJ}(t)+\mathcal{F}^{\mathbb{O} \DD \DD \mathbb{O}}_{JijI}(t)\big]-(t+1)^2&\big[\mathcal{F}^{\mathbb{O} \DD \DD \mathbb{O}}_{IijJ}(t)+\mathcal{F}^{\mathbb{O} \DD \DD \mathbb{O}}_{JjiI}(t)\big]\Big)\log \frac{t}{t+1}\\
&+\Big(t^2\mathcal{F}^{\DD \mathbb{O} \DD \mathbb{O}}_{jIiJ}(t)-\mathcal{F}^{\DD \mathbb{O} \DD \mathbb{O}}_{iIjJ}(t)\Big)\log t\Bigg] =2\(\delta_{ij}\delta_{IJ}\Delta_{\mathbb{O}} + [M_{ij}]_{IJ}\)\,,
\end{split} 
\eeq
and
\beq\la{eqn-Constraint-InHomoFinal-2}
\begin{split}
\Lambda\int\limits_{0}^{\infty} \dd t\,t \Bigg[(t+1)\Big(\mathcal{F}^{\mathbb{O} \DD \DD \mathbb{O}}_{IjiJ}(t) - \mathcal{F}^{\mathbb{O} \DD \DD \mathbb{O}}_{IijJ}(t)+\mathcal{F}^{\mathbb{O} \DD \DD \mathbb{O}}_{JijI}(t)-\mathcal{F}^{\mathbb{O} \DD \DD \mathbb{O}}_{JjiI}(t)&\Big)\log \frac{t}{t+1}\\
&+\(\mathcal{F}^{\DD \mathbb{O} \DD \mathbb{O}}_{iIjJ}(t)- \mathcal{F}^{\DD \mathbb{O} \DD \mathbb{O}}_{jIiJ}(t)\)\log t\Bigg] =[M_{ij}]_{IJ}\,,
\end{split} 
\eeq
\end{widetext}
where $M_{ij}$ is the rotation generator in the $i-j$ transverse plane, acting in the representation of the operator $\mathbb{O}$. For example, in the vector representation, {$[M_{ij}]_{IJ} = \delta_{iJ}\delta_{jI}-\delta_{iI} \delta_{jJ}$. The source on the right-hand side of these constraints originates from the conformal transformation properties of this operator.

Once the defect CFT is specified, the integrated constraints can also be expressed as sum rules on the conformal data. 
These take the schematic form
\beq \label{eqn-SumRule}
\sum_{j}\[A\big(\widetilde\Delta_j\big) C_{\DD \DD\widetilde{\mathbb{O}}_j}C_{\widetilde{\mathbb{O}}_j\mathbb{O} \mathbb{O}} + B\big(\widetilde\Delta_j\big) C^2_{\DD \mathbb{O} \widetilde{\mathbb{O}}_j}\] =\frac{F}{\Lambda}\,,
\eeq
where the sum runs over all DCFT operators and  $A$, $B$, $F$ are universal functions, see Appendix \ref{apd:OPESumRule} for more details.

\section{Checks}\la{checks}

We have performed several checks of the constraints (\ref{eqn-Constraint-HomoFinal}), (\ref{eqn-Constraint-InHomoFinal-1}), (\ref{eqn-Constraint-InHomoFinal-2}) against available data in the literature at weak, strong, and finite coupling. In all cases, we have found them to be satisfied. The examples we considered include 
\begin{itemize}
\item The $1/2$-BPS Wilson line in ${\cal N}=4$ SYM theory in the 't Hooft large $N$ limit. This defect preserves a part of the superconformal symmetry, which relates $\mathbb D$ to other members of a short multiplet. 
We find that the constraints are consistent with these SUSY relations. 
At weak 't Hooft coupling, perturbative data is available up to two loops, \cite{Kiryu:2018phb,Cavaglia:2022qpg}. In the strong coupling limit, the perturbative data (in $1/g$) is available up to four orders \cite{Giombi:2017cqn,Liendo:2018ukf,Ferrero:2023gnu,Ferrero:2023znz}. Finally, the leading OPE behavior of the four-point function is known at finite coupling \cite{Ferrero:2023znz,Ferrero:2023gnu}. 
\item The 1/2-BPS Wilson line in ABJM theory in the ’t Hooft large $N$ limit \cite{Bianchi:2020hsz}. At strong coupling, the displacement four-point function is known at the tree level. It is identical to the result for the $1/2$-BPS Wilson line in ${\cal N}=4$ SYM theory, up to the identification of the string tension with the coupling constant in ABJM theory. Consequently, the checks performed there apply directly to this case.
\item The $1/2$-BPS defect in the CFT dual to string theory on $AdS_3\times S^3\times T^4$ with mixed Ramond-Ramond (RR) and Neveu-Schwarz–Neveu-Schwarz (NS-NS) three-form fluxes \cite{Bliard:2024bcz}. A one-parameter family of such backgrounds, interpolating between the pure NS-NS and pure R-R cases, was studied at strong coupling in \cite{Bliard:2024bcz}. The two-point function and the connected four-point functions of the displacement operator have been computed at leading order. In Appendix \ref{app:ads3}, we demonstrate that in this case the integrated constraints can be employed to fix an undetermined free parameter in the analytic conformal bootstrap.
\item Localized magnetic field line defect in the $O(N)$ Wilson-Fisher fixed point. The two-point function and the connected four-point functions involving at least two displacement operators were calculated in \cite{Cuomo:2021kfm, Gimenez-Grau:2022czc} at the leading order in the $\varepsilon$-expansion.
\end{itemize} 

In the following subsection, we present the check for the localized magnetic field line defect at the $O(N)$ Wilson-Fisher fixed point. This relatively brief example highlights the general features of the weak coupling expansion and allows us to make new predictions.

\subsection{Localized Magnetic Field Line Defect in the $O(N)$ Model}
\label{sec:O(N)}

The $O(N)$ Wilson-Fisher CFT has a simple line defect that was studied in \cite{Cuomo:2021kfm,Gimenez-Grau:2022czc} in the $\varepsilon$-expansion, as we now review. The model consists of $N$ scalars $\vec\phi=(\phi_1,\cdots, \phi_N)$ in $4-\varepsilon$ dimensions, with action
\beq
S = \int\dd^d x\, \Big[ \frac{1}{2} (\partial\vec\phi\,)^2 + \frac{\lambda}{4!} ({\vec\phi}
^{\,2})^2\Big]\,,
\eeq
where, at the fixed point, the coupling constant $\lambda$ takes the value
\begin{equation}
\frac{\lambda_*}{(4\pi)^2} = \frac{3\varepsilon}{N + 8} + \frac{9(3N + 14)\varepsilon^2}{(N + 8)^3} + \mathcal{O}(\varepsilon^3)\,.
\end{equation}

A localized magnetic field defect along a smooth line $\cC$ is introduced by deforming the action as \begin{equation}
S \rightarrow S + h \int\limits_{\mathcal{C}}\, \phi_1\,.
\end{equation}
It breaks the $O(N)$ symmetry down to $O(N-1)$, and induces an RG flow on the defect. The magnetic field $h$ reaches a fixed point at
\begin{equation}
h_*^2 = (N + 8) + \varepsilon \frac{4N^2 + 45N + 170}{2N + 16} + \mathcal{O}(\varepsilon^2)\,.
\end{equation}

The corresponding DCFT contains the following line operators
\begin{description}
\item[$\phi_1$] An $O(N-1)$ singlet of dimension  
$\Delta_{\phi_1} =1  +  \varepsilon + \cO(\varepsilon^2)$.
\item[$\phi_{I}$] A $O(N-1)$ vector of dimension one. These are protected \textit{tilt operators} associated with the broken $O(N)$ symmetry.
\item[$\mathbb{D}$] The displacement operator (\ref{displacement}). It takes the form $\mathbb{D}_j =h_* \partial_j \phi_1$, where $j$ is a transverse vector index.
\end{description}

The relevant conformal data for our computation are
\begin{align} \label{ONdata}
\Delta_{\phi_1} &= 1 + \varepsilon  -\frac{3 N^2+49 N+194}{2 (N+8)^2}\varepsilon^2+ \cO(\varepsilon^3)\,,\nn\\
C_{\DD \DD \phi_1}&= \frac{\pi}{\sqrt{N+8}}\varepsilon +C_{\DD\DD\phi_1}^{(2)}\varepsilon^2+ \cO(\varepsilon^3)\,,\\
C_{\phi_1 \phi_1 \phi_1}&=\frac{3\pi}{\sqrt{N+8}}\varepsilon +C_{\phi_1\phi_1\phi_1}^{(2)}\varepsilon^2+ \cO(\varepsilon^3)\,, \nn\\
\Lambda&= \frac{N+8}{2 \pi ^2}+\frac{5 N^2+23 N+62}{12 \pi ^2 (N+8)}\varepsilon+\cO\left(\varepsilon ^2\right)\,,\nn
\end{align}
where $C_{\DD\DD\phi_1}^{(2)}$ and $C_{\phi_1\phi_1\phi_1}^{(2)}$ are unknown.

In the following, we will focus on the four-point function that involves two displacement operators and two tilt operators or two $O(N-1)$ singlets. These correlators take the form (\ref{eqn-4ptDisDisOO}), (\ref{funnyFs}), with $\mathbb{O}_I = \phi_I$ or $\phi_1$. The tree-level contribution $\cO(\varepsilon^0)$ is a factorized product of two free propagators. In our normalization, this disconnected contribution to $\cF$ does not receive higher-order corrections.

The leading connected contributions $\cO(\varepsilon)$ take the form 
\beq \la{phi1corr}
\mathcal{F}^{\,\phi_1\!\DD\DD\phi_1}_{ij;(1)}(t)=-\frac{\varepsilon\,\delta_{ij} \mathcal{I}_1(t)}{N+8}\,,\ \mathcal{F}^{\,\DD\phi_1\!\DD\phi_1}_{ij;(1)}(t)=-\frac{\varepsilon\,\delta_{ij}\mathcal{I}_2(t)}{N+8}\,,
\eeq
where the subscript $(1)$ indicates the order in the $\varepsilon$-expansion. The two functions of $t$ are given by \cite{Gimenez-Grau:2022czc}
\begin{eqnarray}
\mathcal{I}_1(t) &=&\frac{2 t-1}{t^2}\log (1+t)+\frac{1}{t (t+1)}-\frac{2t+3}{(t+1)^2} \log t\,,\nn\\
\mathcal{I}_2(t) &=& \frac{\log (1+t)}{t^2}-\frac{1}{t (t+1)^2}-\frac{t+3}{(t+1)^3}\log t\,. \label{ONI1I2}
\end{eqnarray}
The four-point functions with the tilt operators are related to these by an overall factor of $\delta_{IJ}/3$. The details of checking the constraints are very similar for that case. Hence, we only present the constraint check with the singlet (\ref{phi1corr}). 

As noted in \cite{Cavaglia:2022qpg}, it is dangerous to substitute perturbative results, such as (\ref{ONI1I2}), directly into finite-coupling integrals. This is because, at some order in the $\varepsilon$-expansion, the corresponding integral may diverge. In such cases, the $\varepsilon$-expansion does not commute with the integration. 

For example, consider an operator $\widetilde{\mathbb{O}}$ of dimension $\widetilde\Delta$ that appears in the $[\DD\times\DD]\to[\mathbb{O}\times\mathbb{O}]$ OPE channel of $\cF^{\mathbb{O}\DD\DD\mathbb{O}}$ in (\ref{eqn-4ptDisDisOO}), (which could be either a primary or a descendant). Its contribution to the four-point function is of the form  
\beq \label{OPEgeneral}
\cF^{\mathbb{O}\DD\DD\mathbb{O}}_{ij}(t)=\delta_{ij}\,\cN\times t^{-\widetilde\Delta}\left(1+\cO(1/t)\right)\,,
\eeq
where $\cN=C_{\DD\DD\widetilde{\mathbb{O}}}C_{\widetilde{\mathbb{O}}\mathbb{O}\mathbb{O}}$ is the product of two structure constants. This behavior results in a contribution to the constraints of the form
\beq\la{regint}
\int\limits^\infty_u\dd t\,t^n\cF^{\mathbb{O}\DD\DD\mathbb{O}}_{11}(t)={\cN\,u^{n+1-\widetilde\Delta}\over\widetilde\Delta-n-1}+\dots\,.
\eeq
where $n<\widetilde\Delta-1$ is an integer and $u$ is a large value for which the OPE expansion (\ref{OPEgeneral}) is still valid.  
Suppose that $\widetilde\Delta = (n+1)+\gamma_1\,\varepsilon+\cO(\varepsilon^2)$, and $\cN=\cN_1\varepsilon+\cN_2\varepsilon^2+\cO(\varepsilon^3)$. 
In perturbation theory, we have
\beq\la{EPOintgeneral}
{\cN\,u^{n+1-\widetilde\Delta}\over\widetilde\Delta-n-1}={\cN_1\over \gamma_1}+\varepsilon{\cN_2 
\over \gamma_1}-\varepsilon\,\cN_1\log u+ \cO(\varepsilon^2)\,.
\eeq
If we naively expand (\ref{OPEgeneral}) in $\varepsilon$ before performing the integration, the $\mathcal{O}(\varepsilon^0)$ term would be missed, and at $\mathcal{O}(\varepsilon^1)$, the integral \eqref{regint} would diverge. This shows that the integral enhances certain contributions in the $\varepsilon$-expansion. Hence, when plugging a perturbative result into the constraints, we must add all the enhanced contributions and regulate the divergences by an analytic continuation in the dimension $\widetilde\Delta$. 

Note that, when $\widetilde\Delta = m+\cO(\varepsilon)$ with $m<n+1$, the integral also diverges for small finite $\varepsilon$. Such divergences are regularized in the standard way, and removed via local counter-terms, leaving a scheme-independent finite result. For example, they can be regulated by analytical continuation in $\widetilde\Delta$, starting from $\widetilde\Delta>n+1$.

\paragraph{The homogeneous constraints}

For the relevant case $i=j$ and $I=J$ in (\ref{eqn-Constraint-HomoFinal}), the first and third lines are equivalent. 
The integrals over $\mathcal{I}_1$ have power divergences. They originate from the $[\phi_1\times\phi_1]\to[\DD\times\DD]$ OPE limit, where
\beq \label{I1larget}
\mathcal{F}^{\,\phi_1\!\DD\DD\phi_1}_{ij;(1)}(t) = -\frac{3 \varepsilon\,\delta_{ij}}{8+N}\(t^{-2}-t^{-3}\)+ \cO(t^{-4})\, .
\eeq
These first two orders match the expansion of the $SL(2,{\mathbb R})$ conformal block of a dimension two operator, see (\ref{conformalblock}). 
There are two such line operators, labeled by their leading-order form as $\phi_1^2$ and $\phi^I\phi_I$. Because $C_{\DD\DD\phi_1^2}C_{\phi_1^2\phi_1\phi_1}=\cO(\varepsilon)$ and $C_{\DD\DD(\phi^I\phi_I)}C_{(\phi^I\phi_I)\phi_1\phi_1}=\cO(\varepsilon^2)$, at order $\cO(\varepsilon)$ only $\phi_1^2$ is exchanged. 
Hence, before integrating $\mathcal{I}_1$ it is sufficient to replace the two leading terms in (\ref{I1larget}) with the finite coupling behavior
\beq\la{cbexp}
t^{-2}\(1-t^{-1}\)\ \rightarrow\ t^{-\Delta_{\phi_1^2}}\Big(1-{\Delta_{\phi_1^2}\over2}t^{-1}\Big)\,.
\eeq
After this substitution the integrals in \eqref{eqn-Constraint-HomoFinal} converge for $\Delta_{\phi_1^2} > 3$, and we re-expand the result in $\varepsilon$. The corresponding enhanced contributions cancel out.

We find that the enhanced contributions for the first homogeneous constraint cancel among themselves. For the second homogeneous constraint, a single enhanced contribution remains. It originates from the exchange of the first two descendants of $\phi_1$ 
in the $[\DD\times\DD] \to [\phi_1\times\phi_1]$ OPE channel 
and is of the form (\ref{EPOintgeneral}) with $\cN_1=0$. 
Collecting all the contributions, we find that the two independent homogeneous constraints in (\ref{eqn-Constraint-HomoFinal}) are satisfied.

\paragraph{The inhomogeneous constraint}

For $I=J$, the second inhomogeneous equation in \eqref{eqn-Constraint-InHomoFinal-2} becomes trivial, so we only have to check the first. 
The source on the right-hand side of this equation 
is of order ${\cO}(\varepsilon^0)$, see (\ref{ONdata}). 
On the left-hand side, the connected four-point function is of order $\mathcal{O}(\varepsilon^1)$, which means that all contributions at order $\mathcal{O}(\varepsilon^0)$ are enhanced ones. They originate from the $\DD$ exchange in the $[\DD\times\phi_1]\to[\DD\times\phi_1]$ and $[\phi_1\times \DD]\to[\DD\times\phi_1]$ OPE channels. Due to a version of \eqref{regint} with logarithms in the kernels, they include enhancements from $\cO(\varepsilon^2)$ to down to $\cO(\varepsilon^0)$. These enhanced contributions sum up to exactly reproduce the right-hand side of the constraint equation.

At next order $\mathcal{O}(\varepsilon^1)$, the inhomogeneous constraint receives both regular and enhanced contributions. 
In this case the enhanced ones arise from the $\phi_1$-exchange in the $[\mathbb{D} \times \mathbb{D}] \to [\phi_1 \times \phi_1]$ OPE channel and the $\mathbb{D}$-exchange in the $[\mathbb{D} \times \phi_1] \to [\mathbb{D} \times \phi_1]$, $[\phi_1 \times \mathbb{D}] \to [\mathbb{D} \times \phi_1]$ OPE channels. 
They combine to an answer that depends on the unknown structure constant $C_{\DD\DD\phi_1}^{(2)}$.  
As a result, the inhomogeneous constraint yields a new prediction
\begin{equation}
\label{eq:CDDphi1}
    C_{\DD\DD\phi_1}^{(2)} = -\frac{\pi}{12}  \frac{29 N^2+413 N+1610}{(N+8)^{5/2}}\,.
\end{equation}

Using a similar analysis, predictions can be extracted for structure constants involving the tilt operator $\phi_I$ and displacements. For example, we find
\begin{equation}
\label{eq:CDDphi2}
    C_{\DD \,\phi_I \,\partial_\perp \phi_I}^{(2)} =\pi \frac{22+N(3-N)}{12 (N+8)^{5/2}}-{\gamma^{(2)}_{\d_\perp \phi_I}\over\sqrt{N+8}}\,.
\end{equation}
}

Further constraints can be formulated using the $\mathcal{O}(\varepsilon^2)$ perturbative data alone. For example, by taking a linear combination of the $\mathcal{O}(\varepsilon^2)$ homogeneous constraints such that the enhanced contribution proportional to $C_{\phi_1 \phi_1 \phi_1}^{(2)}$ drops out, one obtains an integrated constraint on the four-point function at this order that does not involve any other unknown conformal data. Details can be found in Appendix \ref{apd:O(N)eps2}. 

\section{Discussion}\label{sec:discussion}
The expectation value of a critical line defect along an arbitrary smooth path is a conformally invariant functional of its shape. When expanded around a straight path, this invariance imposes an infinite set of integrated constraints on correlation functions involving the displacement operator. 

In this Letter, we have established and studied the first non-trivial constraints in this set; see \eqref{eqn-Constraint-HomoFinal}, \eqref{eqn-Constraint-InHomoFinal-1} and \eqref{eqn-Constraint-InHomoFinal-2}. A comprehensive review of available defect CFT data from the literature revealed that these constraints are universally satisfied in all studied examples. Furthermore, in certain cases, these constraints enabled us to make novel predictions.

The constraints we have considered are only the tip of the iceberg, with many future directions to pursue. Below, we outline several of them: I) The constraints studied in this Letter can be naturally extended by considering higher-order variations. II) In scenarios where the line defect breaks internal symmetry or supersymmetry, protected ``tilt" operators emerge \cite{Bray:1977fvl,Cuomo:2021cnb,Padayasi:2021sik,Herzog:2017xha}. Integrated constraints involving the tilt operators were derived \cite{Behan:2017mwi, Bashmakov:2017rko,Drukker:2022pxk}. These constraints can then be extended to integrated correlators involving both displacement and tilt operator insertions. III) A novel avenue for future exploration is understanding how restrictive these constraints are. For example, in CS-matter theories at large $N$, it has been shown that minimal CFT data, together with these constraints, uniquely determine the line defect CFT and its expectation values \cite{Gabai:2022mya, Gabai:2023lax}. IV) The same ideas may also be implemented for higher-dimensional defects. A new ingredient that appears when an even-dimensional defect is placed along a curved sub-manifold is a gravitational anomaly. Its presence is expected to yield an extra source term in the integrated constraints. V) It would be advantageous to combine these constraints with existing numerical and analytical bootstrap methods to improve our understanding and bounds on defect CFTs.\footnote{Integrated constraints have been proven to be very successful in the CFT bootstrap. One prominent example is the set of constraints derived from supersymmetric localization, \cite{Agmon:2017xes, Binder:2019jwn, Agmon:2019imm, Chester:2021gdw, Alday:2021ymb, Chester:2022sqb, Chester:2021aun, Chester:2023ehi, Chester:2023qwo, Caron-Huot:2024tzr, Caron-Huot:2022sdy, Wen:2022oky, Dorigoni:2022zcr, Dorigoni:2021bvj, Chester:2020dja, Brown:2023zbr, Chester:2020vyz, Alday:2023pet, Dorigoni:2022cua, Paul:2022piq, Dorigoni:2024csx}. Another example involves integrated correlators on supersymmetric $1/2$-BPS Wilson line in $N=4$ super Yang-Mills theory, \cite{Cavaglia:2021bnz, Cavaglia:2022qpg, Cavaglia:2022yvv, Cavaglia:2023mmu,Cavaglia:2024dkk}.} VI) Finally, it would be interesting if similar constraints could also be derived for the correlation functions of the stress-energy tensor of the CFT itself, \cite{inprogress}.\\

{\bf Note}: 
We have been informed that similar lines of investigation have been pursued independently, leading to similar results that will be reported in \cite{Drukker}.

\begin{acknowledgments}
We thank Victor Gorbenko, Shota Komatsu, Ziwen Kong, Slava Rychkov, Philine van Vliet, Miguel Paulos, Michelangelo Preti, and Xi Yin for the discussions. We are grateful to Nikolay Gromov and Zohar Komargodski for insightful discussions and comments on the manuscript. AS is supported by the Israel Science Foundation (grant number 1197/20 and 1099/24). BG is supported by the Simons Collaboration Grant on Confinement and QCD Strings. DlZ is supported in part by the UK Engineering and Physical Sciences Research Council grant number EP/Z000106/1, and the Royal Society under the grant URF\textbackslash R1\textbackslash 221310. 

\end{acknowledgments}

\bibliography{bib}

\newpage
\appendix
\onecolumngrid
\newpage 
\section*{Supplemental Material}

\section{Derivation of the Constraints}\label{derivation app}

The key idea that leads to the integrated constraint is to compute the conformal transformation of a non-straight path in two different ways and then expand 
the results on the straight line. One way is to use the conformal covariance of smooth lines with insertions (or invariance where there are no insertions). That is, consider a spacetime conformal transformation that takes $x\to\tilde x$. Under such a transformation the line defect with insertions transforms as 
\beq\la{source}
{\mathbb O}(\tilde x_1,\tilde n_1){\cal W}[\tilde x(\cdot)]{\mathbb O}(\tilde x_0,\tilde n_0)=[\text{conformal factor}]\times{\mathbb O}(x_1,n_1){\cal W}[x(\cdot)]{\mathbb O}(x_0,n_0)\,,
\eeq
where the $n_i$'s are transverse polarization vectors and the $\tilde n_i$'s are their images under the conformal transformation. Here, we have assumed that that there is no other defect operator that can mix with the operator ${\mathbb O}$. At second order in the deformation, this assumption translates into our second assumption in the main text.

We can further rotate these polarization vectors so that they lie at the intersection of the transverse spaces before and after conformal transformation. Denoting these by $n_i^\text{int}$, equation (\ref{source}) can further be written as
\beq\la{sourcewithspin}
{\mathbb O}(\tilde x_1,n_1^\text{int}){\cal W}[\tilde x(\cdot)]{\mathbb O}(\tilde x_0,n_0^\text{int})=[\text{conformal factor}]\times[\text{spin factor}]\times{\mathbb O}(x_1,n_1^\text{int}){\cal W}[x(\cdot)]{\mathbb O}(x_0,n_0^\text{int})\,.
\eeq

The other way to compute the conformal transformation of the line defect is to express the deformed line as a sum of defect operator insertions on the undeformed line. Requiring consistency between the two approaches leads to a constraint. 
This bootstrap method was first developed to study defects in Chern-Simons matter theory \cite{Gabai:2022vri, Gabai:2022mya, Gabai:2023lax}, but is generally applicable. 

Note that the transformation property (\ref{source}) resembles that of local spacetime {\it primary} operators. In contrast, local descendant operators transform with additional contributions due to mixing with operators involving fewer derivatives. The same applies to defect operators.\footnote{We expect that for defect operators such mixing can be absorbed into the {\it definition} of the operator on a curved line.}
At second order in the deformation, we have found that any such term that is not excluded by the second assumption in the main text does not modify the constraints.

Here, we start from a two-point function on a straight defect. Without loss of generality, we place the two defect operators at the origin and at infinity, $\llangle \mathbb{O}(0) \mathbb{O}(\infty) \rrangle$. We then smoothly deform the path as in (\ref{deformation}). For the order in which we work in this paper, it is sufficient to focus on the second order in the deformation parameter, $\v$. 
The terms at this order can be classified according to the number of operators that are been integrated on the line
\beq\la{eq:2variation0}
\delta^2\llangle \mathbb{O}(0) \mathbb{O}(\infty) \rrangle=[{\rm int^2}] + [{\rm int}] + [B^2_0] + [B^2_{\infty}] + [B_0B_{\infty}]\,.
\eeq
Here, $[{\rm int^2}]$ is a double integral over two displacement operators, $[{\rm int}]$ is a single integral, and $[B^2]$, $[B^2_{\infty}]$, $[B_0B_{\infty}]$ are boundary terms, without integrations. They consist of all the operators that are allowed by the symmetries of the straight line and are listed in the following sections.

We then decompose the transverse deformation vector in (\ref{deformation}) as $\v=\v_a+\v_c$, where the transverse deformation vector $\v_c$ corresponds to a conformal deformation and $\v_a$ is an arbitrary smooth deformation that is not associated with a conformal one. These two are independent at the second order that we are working at.

By equating (\ref{eq:2variation0}) with the conformal transformation of the two operators at orders $\mathcal{O}(\v_c^2)$ and $\mathcal{O}(\v_a \v_c)$ we can fix all coefficients in front of the operators appearing in \eqref{eq:2variation0}. We now describe each of the terms in detail and then derive the constraints from them. 

\subsection{Double Integration} 

The double integration term consists of two ordered insertions of displacement operators along the line. They come in three possible orderings
\begin{align}\la{eqn-2ndOrd-Int2}
[{\mathrm{int}^2}] =& \int\limits_{\tau_1<0<\tau_2}\!\!\!\!\!\!\dd \tau_1 \dd\tau_2\,\v^i(\tau_1)\v^j(\tau_2)\,\llangle \mathbb{D}_i(\tau_1) \mathbb{O}_I(0) \mathbb{D}_j(\tau_2) \mathbb{O}_J(\infty) \rrangle \nn\\ +& \int\limits_{0<\tau_1<\tau_2}\!\!\!\!\!\!\dd \tau_1 \dd \tau_2\,\v^i(\tau_1)\v^j(\tau_2)\,\llangle\mathbb{O}_I(0)\mathbb{D}_i(\tau_1)\mathbb{D}_j(\tau_2)\mathbb{O}_J(\infty) \rrangle\\
+&\int\limits_{\tau_2<\tau_1<0}\!\!\!\!\!\!\dd \tau_1 \dd \tau_2\,\v^i(\tau_1)\v^j(\tau_2)\,\llangle\mathbb{D}_j(\tau_2)\mathbb{D}_i(\tau_1)\mathbb{O}_I(0)\mathbb{O}_J(\infty) \rrangle\,,\nn
\end{align}
where we are using a ``proper time" parametrization of the straight line in which $x(\tau)=\tau$, so $|\d_\tau x(\tau)|=1$ in (\ref{displacement}). These integrals are generically divergent. We regularize them using point splitting in a conformal frame where the two $\mathbb O$ operators are at $\tau=0$ and $\tau=1$. After using the conformal transformation $\tau\to \tau/(1-\tau)$ to map this frame to the one where the operators are inserted at $\tau=0$ and $\tau=\infty$, this regularization scheme reads
\beq\la{pointsplittingreg}
\eps<|\tau_j|<1/\eps\,,\qquad {|\tau_2-\tau_1|}>\epsilon|(1+\tau_1)(1+\tau_2)|\,.
\eeq

Using the normalization where $\llangle{\mathbb O}_I(0){\mathbb O}_J(x)\rrangle=\delta_{IJ}/|x|^{2\Delta}$, the correlators in $[\text{int}^2]$ are related to the ones in (\ref{eqn-4ptDisDisOO}) and (\ref{funnyFs}) as
\beq
\begin{aligned}
\llangle \mathbb{D}_i(x_1 ) \mathbb{O}_I(x_2) \mathbb{D}_j(x_3 ) \mathbb{O}_J(\infty) \rrangle & = \frac{\Lambda}{x_{23}^4} \times  \widehat{\mathcal{F}}_{iIjJ}^{\mathbb{D}\mathbb{O}\mathbb{D}\mathbb{O}}(t) \, , \\  
\llangle \mathbb{O}_I(x_1) \mathbb{D}_i(x_2 )  \mathbb{D}_j(x_3 ) \mathbb{O}_J(\infty)  \rrangle & = \frac{\Lambda}{x_{23}^4 }   \times \widehat{\mathcal{F}}_{IijJ}^{\mathbb{O}\mathbb{D}\mathbb{D}\mathbb{O}}(t)\, , \\  
\llangle \mathbb{D}_j(x_1 )  \mathbb{D}_i(x_2 ) \mathbb{O}_I(x_3) \mathbb{O}_J(\infty)  \rrangle & = \frac{\Lambda}{x_{12}^4 }   \times \widehat{\mathcal{F}}_{JjiI}^{\mathbb{O}\mathbb{D}\mathbb{D}\mathbb{O}}\(1/t\)\, .
\end{aligned}
\eeq
These are defined so that for ${\mathbb O}={\mathbb D}$ they are all equal. 
In terms of these $\widehat{\cF}$'s, \eqref{eqn-2ndOrd-Int2} takes the form
\begin{align}\la{eq:2variationTilde}
[{\rm int^2}]/\Lambda=&\quad\int\limits_{0}^{\infty}\quad\ \,\frac{\dd \tau_1 \dd \tau_2}{\tau_{2}^4} \widehat{\mathcal{F}}_{iIjJ}^{\mathbb{D}\mathbb{O}\mathbb{D}\mathbb{O}}\left(\tau_1/\tau_2\right) \v^i(-\tau_1)\v^j(\tau_2)\\
+&\!\int\limits_{0<\tau_1<\tau_2} \frac{\dd \tau_1 \dd \tau_2}{(\tau_2-\tau_1)^4}\[\widehat{\mathcal{F}}_{IijJ}^{\mathbb{O}\mathbb{D}\mathbb{D}\mathbb{O}}\Big(\frac{\tau_1}{\tau_2-\tau_1}\Big)\v^i(\tau_1)\v^i(\tau_2)+\widehat{\mathcal{F}}_{JjiI}^{\mathbb{O}\mathbb{D}\mathbb{D}\mathbb{O}}\Big(\frac{\tau_1}{\tau_2-\tau_1}\Big)\v^i(-\tau_1)\v^j(-\tau_2)\]\,,\nn
\end{align}  
where we leave the point-splitting regularization implicit. Here, we have made some changes of the integration variables that is inferred from the argument of the $\v$'s. 

Next, we rewrite the double integral (\ref{eq:2variationTilde}) in terms of the subtracted correlators $\mathcal{F} = \widehat{\mathcal{F}}- \mathcal{F}_\text{GFF}$, where $\cF_\text{GFF}$ is given in (\ref{eq:GFFpiece}). We find that $[\text{int}^2]$ takes the same form in terms of the subtracted correlators. To see that, 
we plug $\cF_\text{GFF}$ into \eqref{eq:2variationTilde} and take the deformation vector $\v$ to be a combination of the deformations that we will use in the following to derive the constraints (see (\ref{vc}) and (\ref{va})). We then perform the integrations using the regularization scheme (\ref{pointsplittingreg}). After expanding the result in $\epsilon$, we find that there are no finite contributions remaining. 

The integrals in (\ref{eq:2variationTilde}) have a convolution structure. Hence, it is natural to express them in Mellin space,
\beq\la{eqn-VLaplaceConvention}
\tilde{f}(s) = \int\limits_0^\infty \dd\tau\, \tau^{s-1}f(\tau)\,, \qquad f(\tau) = \int\limits_{c-\ii\infty}^{c+\ii\infty} \frac{\dd s}{2\pi \ii} \ \tau^{-s} \tilde{f}(s)\ ,\  
\eeq
where the real part of the Mellin contour $c$  is determined by the small and large $\tau$  asymptotics of the function $f(\tau)$.

For pure Mellin modes $\v^i(\tau_1)=\tau_1^{-s_1}$ and $\v^j(\tau_2)=\tau_2^{-s_2}$, the first line in $[\text{int}^2]$, (\ref{eq:2variationTilde}) can be written in terms of the Mellin transformed correlator as
\begin{equation}\label{eq:2intF}
\begin{split}
&\int\limits_{\epsilon}^{1/\epsilon} \frac{\dd \tau_1 \dd \tau_2}{\tau_2^4} \tau_1^{-s_1} \tau_2^{-s_2} \mathcal{F}_{iIjJ}^{\mathbb{D}\mathbb{O}\mathbb{D}\mathbb{O}} \(\tau_1/\tau_2\) =  \int\limits_{\epsilon}^{1/\epsilon} \dd \tau_2\int\limits_{\epsilon/\tau_2}^{1/(\tau_2\epsilon)}\!\! \dd t \, \tau_2^{-s_2-s_1-3} t^{-s_1}\mathcal{F}_{iIjJ}^{\mathbb{D}\mathbb{O}\mathbb{D}\mathbb{O}}(t) \\
    &=  \int\limits_{\epsilon}^{1/\epsilon} \dd \tau_2\int\limits_{0}^{\infty} \dd t \, \tau_2^{-s_2-s_1-3} t^{-s_1}\mathcal{F}_{iIjJ}^{\mathbb{D}\mathbb{O}\mathbb{D}\mathbb{O}}(t)-  \int\limits_{\epsilon}^{1/\epsilon} \dd \tau_2\Bigg[\int\limits_{0}^{\eps/\tau_2}+\int\limits_{1/(\eps \tau_2)}^{\infty}\Bigg] \dd t \, \tau_2^{-s_2-s_1-3} t^{-s_1}\mathcal{F}_{iIjJ}^{\mathbb{D}\mathbb{O}\mathbb{D}\mathbb{O}}(t) \\
    & = 2\pi \ii \delta_\epsilon(s_1+s_2+2) \widetilde{\mathcal{F}}^{\mathbb{D}\mathbb{O}\mathbb{D}\mathbb{O}}_{iIjJ}(1-s_1) + [\text{boundary-contributions}]\,.
\end{split}
\end{equation}
where $[\text{boundary-contributions}]$ stands for terms in which at least one of the integrations is localized near the boundary. Such terms are absorbed in the boundary and single integration terms in (\ref{eq:2variation0}) that are analyzed in the next subsections. 
Here, $\delta_\epsilon$ is the regularized delta function
\begin{equation}
    \delta_\epsilon(x)\equiv \int\limits_{\epsilon}^{1/\epsilon} \frac{d\tau}{\tau}\tau^x \,.
\end{equation}
Similarly, the integrals on the second line of (\ref{eq:2variationTilde}) can be evaluated using
\begin{align}\label{eq:2intH}
\int\limits_{0<\tau_1<\tau_2} \frac{\dd \tau_1 \dd \tau_2}{(\tau_2-\tau_1)^4} \tau_1^{-s_1} \tau_2^{-s_2} \mathcal{F}^{\mathbb{O}\mathbb{D}\mathbb{D}\mathbb{O}}_{IijJ} \(\frac{\tau_1}{\tau_2-\tau_1}\) &=  \int\limits_{2\eps}^{1/\eps} \dd \tau_2 \,\tau_2^{-s_1-s_2-3}\!\!\!\!\int\limits_{\eps/\tau_2}^{1-\frac{\eps}{\tau_2(1+\tau_2)^2}}\!\!\!\!\dd x\  \frac{x^{-s_1}}{(x-1)^4} \mathcal{F}_{IijJ}^{\mathbb{O}\mathbb{D}\mathbb{D}\mathbb{O}}\(\frac{x}{1-x}\)\\ 
&= 2\pi \ii\delta_\epsilon(s_1+s_2+2) \widetilde{\mathcal{H}}^{\mathbb{O}\mathbb{D}\mathbb{D}\mathbb{O}}_{IijJ}(1-s_1) + [\text{boundary-contributions}]\,,\nn
\end{align}
where 
\beq\la{Htilde}
\widetilde{\mathcal{H}}^{\mathbb{O}\mathbb{D}\mathbb{D}\mathbb{O}}_{IijJ}(s) \equiv \int\limits_0^\infty {\dd t}\,{(1 + t)^2} \(\frac{t}{t+1}\)^{s-1}\!\!\mathcal{F}_{IijJ}^{\mathbb{O}\mathbb{D}\mathbb{D}\mathbb{O}}\(t\) \,.
\eeq
By introducing the inverse Mellin integrals
\beq\la{eqn-VLaplaceConvention}
\tilde{\v}_{\pm}(s)\equiv\int\limits_0^\infty \dd\tau\,\tau^{s-1}\,\v(\pm\tau)\,, \qquad \v(\pm|\tau|) = \int\limits_{c-\ii\infty}^{c+\ii\infty} \frac{\dd s}{2\pi \ii} \ \tau^{-s}\,\tilde{\v}_\pm(s)\,,
\eeq
we arrive at
\begin{align}\la{eq:afterMellin}
[{\rm int^2}]=\Lambda\int \frac{\dd s_1 \dd s_2}{2\pi \ii}&\delta_\eps(2+s_1+s_2)\\
\times&\Big[ \widetilde{\mathcal{F}}^{\mathbb{D}\mathbb{O}\mathbb{D}\mathbb{O}}_{iIjJ}(1-s_1) \tilde{\v}^i_{-}(s_1) \tilde{\v}_{+}^j(s_2)+\widetilde{\mathcal{H}}^{\mathbb{O}\mathbb{D}\mathbb{D}\mathbb{O}}_{IijJ}(1-s_1) \tilde{\v}^i_{+}(s_1) \tilde{\v}_{+}^j(s_2)+\widetilde{\mathcal{H}}^{\mathbb{O}\mathbb{D}\mathbb{D}\mathbb{O}}_{JjiI}(1-s_1) \tilde{\v}^i_{-}(s_1) \tilde{\v}_{-}^j(s_2) \Big]\,.\nn
\end{align}  

\subsection{Single Integration}
\label{sec:singint}

For a single integral, the two deformations can be either integrated together or one is integrated and the other is at the location of one of the $\mathbb O$ operators. We would like to enumerate all such terms that are consistent with the $SL(2,{\mathbb R}\times SO(d-1)$ symmetry of the straight line. This symmetry imposes two conditions. First, the boundary terms must retain the same $SL(2,{\mathbb R})$ dimension as $\mathbb O$ prior to the deformation; second, the integrand on the defect must have dimension one, ensuring that the integral remains dimensionless. We assume that the CFT is ``generic". That is, we assume that 
\begin{enumerate}
\item The only scalar operator that is a singlet under all internal symmetries and has integer dimension $\Delta<4$ is the identity.
\item There is no operator of dimension $\Delta_{\mathbb O}+1$ that otherwise has the same charges as $\mathbb O$ and a spin larger by one unit.
\end{enumerate}

Under this assumption, we find that for a generic spacetime dimension, the only terms with a single bulk integration and a non-zero expectation value are total derivatives. Such terms can be absorbed in the boundary terms in (\ref{eq:2variation0}). For example, we have
\begin{equation}
\int\limits_a^b \dd\tau\, \v(\tau) \v'''(\tau) \times\One=\left.\big[\v(\tau) \v''(\tau) - \frac{1}{2} \left(\v'(\tau)\right)^2\big]\right|_a^b\,.
\eeq

In two, three, and four spacetime dimensions an additional operator on the line is allowed. Specifically, in four spacetime dimensions the following single integral is consistent with the straight-line symmetries
\beq
\int\dd\tau\,\epsilon^{ijk}\v^i(\tau){\v'}^j(\tau)\llangle{\mathbb O}_I(0)\DD^k(\tau){\mathbb O}_J(\infty)\,\rrangle\in[\text{int}]_{4d}\,,
\eeq
where the $i$-index runs over the three transverse directions. 
We have found, however, that this term is ruled out by our bootstrap (although we do not include the detailed analysis in this note). Hence, we will not include it in the following.
    
Similarly, in three spacetime dimensions the following term is allowed
\beq
\int\dd\tau\,\epsilon^{ij}\v^i(\tau){\v'}^j(\tau)\llangle{\mathbb O}_I(0){\mathbb O}_J(\infty)\,\rrangle\in[\text{int}]_{3d}\,.
\eeq
While this term is not a total derivative, its addition has the effect of shifting the spin source on the right-hand side of the inhomogeneous constraints (\ref{eqn-Constraint-InHomoFinal-1}), (\ref{eqn-Constraint-InHomoFinal-2}).  
In our earlier work on Chern-Simons matter theories \cite{Gabai:2022vri,Gabai:2023lax}, this term was excluded at third order, $O(\v^3)$, of the bootstrap. This fact implies that also for any other three-dimensional theory, the coefficient of this term is fixed at third order. We expected, but did not prove, that it would be fixed to zero. In what follows we assume that this is indeed the case and therefore the constraints are not altered as we go down to three spacetime dimensions.

In two-dimensional CFTs in which there is no reflection symmetry in the direction perpendicular to the line the integral 
\beq
\int\dd\tau\,\v(\tau)\dd {\v}(\tau)\llangle{\mathbb O}_I(0)\DD (\tau){\mathbb O}_J(\infty)\,\rrangle\in[\text{int}]_{2d}\,,
\eeq
is allowed. Similarly, there are also new boundary terms that can be added. In this case, we have repeated the derivation of the constraints for the four-point function of the displacement operator and found that they are not modified. This fact will be relevant for us when checking the constraints in the two-dimensional example studied in appendix \ref{app:ads3}.

\subsection{Boundary Terms}

We now list all the boundary terms that are allowed by the $SL(2,{\mathbb R})\times SO(d-1)$ symmetry of the straight line. In total, such terms have zero $SO(d-1)$ transverse spin and zero $SL(2,{\mathbb R})$ conformal dimension. Since the two $\v$'s together have dimension minus two and spin zero, these terms consist of the operator $\mathbb O$, its first and second descendants.

We take the straight line to be placed along the third direction, $x_3$. For simplicity, we restrict our discussion to the case where $\v$ is in a two-dimensional transverse space that is spanned by $x_\pm=(x_1\pm\ii x_2)/\sqrt 2$. This restricted kinematical configuration is sufficient for the derivation of the constraints. The boundary terms that are allowed in any CFT are 
\beqa
[B_0B_0]&=&\llangle\big[ \gamma_0\v_0^+ \v_0^-\dd^2\mathbb{O}_I(0)+\gamma_1\(\v_0^+ \dd \v_0^- + \v_0^- \dd \v_0^+\)\dd\mathbb{O}_I(0)+\gamma_4\(\v_0^+ \dd \v_0^- - \v_0^- \dd \v_0^+\)\dd\mathbb{O}_I(0)\nn\\
&&+\(\gamma_3 \dd \v_0^+ \dd \v_0^-+\gamma_2 \(\v_0^+ \dd \dd \v_0^- + \v_0^- \dd \dd \v_0^+\)+\gamma_5 \(\v_0^+ \dd \dd \v_0^- - \v_0^- \dd \dd \v_0^+\)\)\mathbb{O}_I(0)\big]  \mathbb{O}_J(\infty)  \rrangle\,,\la{eq:BB}\\
\,[B_\infty B_\infty]&=&\llangle \mathbb{O}_I(0)\big[\tilde\gamma_0\v_1^+ \v_1^-\dd^2\mathbb{O}_J(\infty) +\tilde\gamma_1\(\v_\infty^+ \dd \v_\infty^- + \v_\infty^- \dd \v_\infty^+\)\dd\mathbb{O}_J(\infty) -\tilde\gamma_4\(\v_\infty^+ \dd \v_\infty^- - \v_\infty^- \dd \v_\infty^+\)\dd\mathbb{O}_J(\infty)\nn\\
&&+\(\tilde\gamma_3 \dd \v_\infty^+ \dd \v_\infty^-+\tilde\gamma_2 \(\v_\infty^+ \dd \dd \v_\infty^- + \v_\infty^- \dd \dd \v_\infty^+\) - \tilde\gamma_5 \(\v_\infty^+ \dd \dd \v_\infty^- - \v_\infty^- \dd \dd \v_\infty^+\)\)\mathbb{O}_J(\infty)\big] \rrangle\,,\nn
\eeqa
where 
$\gamma_i$, $\tilde{\gamma}_i$ are unknown coefficients. Here, $\dd\v$ and $\dd{\mathbb O}$ stand for a derivative of $\v$ and ${\mathbb O}$ along the straight line. As for descendant operators, the form of these derivatives depends on the conformal frame in which they are defined, which is tailored to the scheme in which we regulate the line integrals. Recall that we use the point-splitting regularization scheme in the conformal frame where the two operators are placed at $\tau=0$ and $\tau=1$. Hence, in this frame $\dd\v=\d_\tau\v$ and $\dd{\mathbb O}=\d_\tau{\mathbb O}$, where $x^3(\tau)=\tau$ with $\tau\in[0,1]$. Using the conformal transformation $w\to w/(1-w)$ employed in (\ref{pointsplittingreg}), this definition translates to
\beq
\label{eq:defvbdry}
\dd^n A(\tau) \equiv \lim_{\sigma\to\tau}\((1+\sigma)^{2}\d_\sigma\)^n \frac{A(\sigma)}{(1+\sigma)^2}\,,
\eeq
the $[0,\infty]$ frame.

Additionally, there are terms that are allowed when the defect contains operators of special dimension and spin. For example
\begin{itemize}
\item If the theory has an operator $\widetilde{\mathbb{O}}$ of dimensions $\Delta_{\widetilde{\mathbb{O}}}=\Delta_{\mathbb{O}} - m$ where $m$ is a positive integer, then there may be additional allowed boundary terms. For example, if $\mathbb{O} =\DD$ and $\widetilde{\mathbb O}=\One$ then the term
\begin{equation}
\label{eq:B0BinfId}
\dd\dd\dd \v^i_0\, \dd\dd\dd \v^j_\infty\ \in\ \delta^2\llangle\mathbb{D}^i(0) \mathbb{D}^j(\infty)\rrangle\,,
\end{equation}
is allowed and is part of $[B_0B_{\infty}]$ in the decomposition (\ref{eq:2variation0}). This term, however, does not contribute when at least one of the $\v$'s corresponds to a conformal transformation, which is the case relevant to our discussion. When $\widetilde{\mathbb O}\neq \One$, these can be interpreted as the operator mixing mentioned below (\ref{sourcewithspin}) and above \eqref{eqn-Constraint-HomoFinal}.
\item If $\mathbb{O}$ has transverse spin $1/2$ then terms such as 
\beq
\dd \v^\pm_0\, \dd \v^\mp_\infty\llangle\mathbb{O}_{\pm{1\over2}} \mathbb{O}_{\mp{1\over2}}\rrangle\ \in\ \delta^2\llangle\mathbb{O}_{\mp{1\over2}}(0) \mathbb{O}_{\pm{1\over2}}\rrangle\,,
\eeq
are allowed and are part of $[B_0B_{\infty}]$ in the decomposition (\ref{eq:2variation0}). 
\item If $\mathbb{O}$ of spin one, then terms such as
\beq
\v_0^+ \dd \v_0^+\llangle \dd\mathbb{O}_+(0)  \mathbb{O}_-(\infty)\rrangle\ \in\ \delta^2\llangle \mathbb{O}_-(0)\mathbb{O}_-(\infty)\rrangle\,,
\eeq
are allowed. In this case, the two $\v$'s together have transverse spin 2.
\end{itemize}
Apart from the cases that are excluded by our assumptions in the main text, in all these instances, we have verified that the presence of the new terms does not affect the constraints. To streamline the presentation, 
in the following we will assume that they are not present.

\subsection{Source Term}

We now choose the deformation vector in (\ref{deformation}) that will lead to the constraints presented in the main text. We start with a non-straight path parameterized by $\v(\tau)=\v_a(\tau)$. At the second order, applying to it a conformal transformation amounts to simply adding to it the infinitesimal conformal variation vector, $\v_a(\tau)\to\v_a(\tau)+\v_c(\tau)$.\footnote{At higher orders $\v_c$ depends on $\v_a$.} 
Consequently, the change in the expectation value of the deformed line receives contributions at orders $\mathcal{O}(\v_c^2)$ and $\mathcal{O}(\v_c \v_a)$. The source term of $\delta^2 \langle \mathbb{O}(0) \mathbb{O}(\infty) \rangle$ has two contributions. The first arises from the dilatation of the two operators, and the second arises from their transverse rotation, which are denoted by $[\mathtt{conformal \ factor}]$ and $[\mathtt{spin \ factor}]$ in (\ref{sourcewithspin}).
The derivation of these contributions in the three spacetime dimensions can be found in section 2 of \cite{Gabai:2023lax}, and equally applies to higher dimensions. In more details, any conformal transformation can be decomposed into a translation, a rotation, and a dilatation,
\beq
\frac{\partial \tilde{x}^\mu}{\partial x^\nu} = \Omega(x) {\Lambda^\mu}_\nu(x)\,, \qquad \text{where} \quad \det {\Lambda^\mu}_\nu = 1\, .
\eeq
The translation and rotation of the defect operator together with the line and its polarization vector do not induce a transformation of the operator. In contrast, dilation is part of the $SL(2,{\mathbb R})$ subgroup, in which the boundary operator has a well-defined conformal weight. At second order it takes the form
\beq\label{eqn-2ndOrd-Source}
[\mathtt{conformal \ factor}]=-\Delta\,\delta_{IJ}(\vec\v_\infty-\vec\v_0)^2\,.
\eeq

Next, the rotation of the polarizations vectors in (\ref{sourcewithspin}) to the intersection of the transverse spaces before and after the conformal transformation leads to the additional factor
\begin{align}
[\mathtt{spin\ source}]_{\cO(\v_c^2)} =&\, \frac{1}{2}\left.\Big[\[M_{ij}\]_{IJ} \v_c^i \dd \dd \v_c^j /2\Big]\right|^{\infty}_0\,,
\la{eq:spin_vcsq}\\
[{\mathtt{spin\ source}}]_{\cO(\v_c \v_a)} =& \left.\Big[\[M_{ij}\]_{IJ}\(\v_a^i \dd \dd \v_c^j + \dd\v_a^i \dd \v_c^j/2\)\Big]\right|^{\infty}_0\,,\la{eq:spin_vcva}
\end{align}
where $M_{ij}$ is the rotation generator in the $i-j$ transverse plane, acting in the representation of the operator $\mathbb{O}$. For example, in the vector representation, $[M_{ij}]_{IJ} = \delta_{iJ}\delta_{jI}-\delta_{iI} \delta_{jJ}$.

Finally, an infinitesimal conformal transformation is a degree two polynomials with arbitrary vector coefficients that we denote by $\mathfrak{c}$,
\beq \la{vc}
\v_c^{i}(\tau) = \mathfrak{c}^i_0 + \mathfrak{c}^i_{1}\tau+ \mathfrak{c}^i_{2}\tau^2\,.
\eeq
The Mellin transform (\ref{eqn-VLaplaceConvention}) of $\v_c(\tau)$ is
\beq\la{vconformal}
    \tilde{\v}^i_{c,\pm}(s) = 2\pi \ii \[\mathfrak{c}^i_0\delta(s) \pm \mathfrak{c}^i_1 \delta(1+s) + \mathfrak{c}^i_2 \delta(2+s)\] \,.
\eeq

\subsection{Constraints}

The constraints arise from equating $\delta^2\llangle \mathbb{O}(0) \mathbb{O}(\infty) \rrangle$ in (\ref{eq:2variation0}) with the source term at orders $O(\v_c^2)$ and $O(\v_a\v_c)$. We consider each of these orders in turn and separate between the cases where $\tilde\v_a(s)$ is regular or has a pole at the support of the delta function in $[\text{int}^2]$, (\ref{eq:afterMellin}). 

\subsubsection{$O(\v_a \v_c)$ Constraints from Regular $\v_a$}
We first consider profiles $\v_a(\tau)$ that vanish sufficiently fast at the boundaries, such that the boundary terms $[B_0B_0]$ and $[B_\infty B_\infty]$ in \eqref{eq:BB} vanish. For such a choice, the source terms in (\ref{eqn-2ndOrd-Source})-(\ref{eq:spin_vcva}) also vanish. Moreover, for ${\mathbb O}={\mathbb D}$, the term in (\ref{eq:B0BinfId}) 
vanishes because $\v_c$ is a degree $2$ polynomial. Hence, the only contribution with which we remain is the double integral $[\text{int}^2]$ and the corresponding constraint is homogeneous. 

With this choice, the Mellin transformed $\tilde{\v}^i_a(s)$ is regular at the support of the delta function in \eqref{eq:afterMellin}, $s=0,-1,-2$, so we can safely take the $\epsilon \rightarrow 0$ limit there. Plugging in $\tilde\v_c$ from (\ref{vconformal}) leads to the expression
\begin{equation} \label{eqn-2ndOrd-DoubleInt-vavc-nonsing}
\begin{split}
    [{\rm int^2}]\Big|_{\mathcal{O}(\v_a \v_c)} = \sum_{\ell=0}^{2} \mathfrak{c}^i_\ell \Big(&\tilde{\v}^j_{a,-}(\ell-2)\big[\widetilde{\mathcal{F}}^{\mathbb{D}\mathbb{O}\mathbb{D}\mathbb{O}}_{jIiJ}(3-\ell) +(-1)^\ell\widetilde{\mathcal{H}}^{\mathbb{O}\mathbb{D}\mathbb{D}\mathbb{O}}_{JijI}(3-\ell)+(-1)^\ell\widetilde{{H}}^{\mathbb{O}\mathbb{D}\mathbb{D}\mathbb{O}}_{JjiI}(\ell+1)\big]\\ 
    +& \tilde{\v}^j_{a,+}(\ell-2)\big[(-1)^\ell\widetilde{\mathcal{F}}^{\mathbb{D}\mathbb{O}\mathbb{D}\mathbb{O}}_{iIjJ}(\ell+1)+\widetilde{\mathcal{H}}^{\mathbb{O}\mathbb{D}\mathbb{D}\mathbb{O}}_{IjiJ}(3-\ell)+\widetilde{{H}}^{\mathbb{O}\mathbb{D}\mathbb{D}\mathbb{O}}_{IijJ}(\ell+1)\big]\Big)\,.
\end{split}
\end{equation}
Since 
$\tilde{v}_{a,\pm}(\ell-2)$ are free parameters, the equation imposes the following homogeneous constraints
\begin{equation}
\begin{split}\label{eq:HomoConstDer}
    (-1)^\ell \widetilde{\mathcal{F}}^{\mathbb{D}\mathbb{O}\mathbb{D}\mathbb{O}}_{jIiJ}(3-\ell) +\widetilde{\mathcal{H}}^{\mathbb{O}\mathbb{D}\mathbb{D}\mathbb{O}}_{JijI}(3-\ell)+\widetilde{{H}}^{\mathbb{O}\mathbb{D}\mathbb{D}\mathbb{O}}_{JjiI}(1+\ell) &= 0 \\
    (-1)^\ell  \widetilde{\mathcal{F}}^{\mathbb{D}\mathbb{O}\mathbb{D}\mathbb{O}}_{iIjJ}(1+\ell) + \widetilde{\mathcal{H}}^{\mathbb{O}\mathbb{D}\mathbb{D}\mathbb{O}}_{IjiJ}(3-\ell)+\widetilde{{H}}^{\mathbb{O}\mathbb{D}\mathbb{D}\mathbb{O}}_{IijJ}(1+\ell) &= 0
\end{split}
\,,\qquad\ell=0,1,2\,.
\end{equation}
These constraints apply for any choice of transverse directions $i$ and $j$. Only three out of these six constraints are independent. Specifically, the first and the second lines are interchanged under $i\leftrightarrow j$, $I\leftrightarrow J$, and $\ell\leftrightarrow2-\ell$. After plugging in the definitions of the Mellin transforms $\widetilde\cF$ (\ref{eqn-VLaplaceConvention}) and $\widetilde{\cal H}$ (\ref{Htilde}), we arrive at the homogeneous constraints (\ref{eqn-Constraint-HomoFinal}) in the main text.

\subsubsection{$\cO(v_c^2)$ Constraints}

Consider first the double integral at order $O(\v_c^2)$. By 
plugging $\tilde\v_c$ in \eqref{vconformal} into \eqref{eq:afterMellin} we arrive at 
\beq
[{\rm int^2}]\Big|_{\mathcal{O}(\v_c^2)}=2\pi \ii\,\Lambda\sum_{\ell_1,\ell_2=0}^2 \mathfrak{c}_{\ell_1}^i \mathfrak{c}_{\ell_2}^j \delta_\eps(2-\ell_1-\ell_2)\Big[ (-1)^{\ell_1}\widetilde{\mathcal{F}}^{\mathbb{D}\mathbb{O}\mathbb{D}\mathbb{O}}_{iIjJ}(1+\ell_1)+\widetilde{\mathcal{H}}^{\mathbb{O}\mathbb{D}\mathbb{D}\mathbb{O}}_{IijJ}(1-\ell_1) + (-1)^{\ell_1+\ell_2}\widetilde{\mathcal{H}}^{\mathbb{O}\mathbb{D}\mathbb{D}\mathbb{O}}_{JjiI}(1-\ell_1)  \Big]\,.
\eeq
The combination in the parentheses vanishes separately for each value of $\ell_1$ and $\ell_2$ by \eqref{eq:HomoConstDer}.

We remained with the boundary and source terms. By comparing them for $\mathfrak{c}_i$'s in a two-dimensional transverse plane we arrive at the following equations for the boundary coefficients 
\beq \label{eq:gammasConstrained}
\begin{aligned}
\tilde\gamma_0 = &-\gamma_0 = 
\frac{1-2\gamma_1}{1+2\Delta}\,,\qquad
\gamma_1=1-\tilde\gamma_1=\frac{\gamma_2+\gamma_3}{\Delta}\,,\qquad
\tilde\gamma_2+\gamma_2= \Delta\,,\\
\gamma_3=&\tilde\gamma_3\,,\qquad
\gamma_4=\tilde\gamma_4\,,\qquad\Delta \gamma_4 - \gamma_5=\mathfrak{s}_0/2\,,\qquad\Delta\tilde\gamma_4 + \tilde\gamma_5 =\mathfrak{s}_\infty/2 \,,
\end{aligned}
\eeq
where $\mathfrak{s}_0$ and $\mathfrak{s}_\infty$ are the clockwise $U(1)$ spins of the two operators in the $x_1$--$x_2$ transverse plane that we use here. This implies that $\gamma_5$ and $\tilde{\gamma}_5$ depend on the polarizations of the two ${\mathbb O}$ operators. 

\subsubsection{$O(\v_a \v_c)$ Constraints from More General $\v_a$}

We consider $\v_a$ profiles of the form
\beq\la{va}
    \v^i_a \in \Big\{\frac{\tau}{\tau^2+1}\,,\ \frac{\tau^2}{\tau^2+1}\,,\ \frac{\tau^3}{\tau^2+1}\Big\}\,.
\eeq
Their Mellin transform, $\tilde{\v}_a^i(s)$, has a simple pole at the support of the delta function in \eqref{eq:afterMellin}. That is, at $s=0$, $s=-1$ or $s=-2$. However, these poles multiply expressions that vanish by the homophonous constraint \eqref{eq:HomoConstDer}. As a result, the corresponding integrand in $[\text{int}^2]$ is regular at $\epsilon\to0$. Yet, to obtain the finite terms in $[\text{int}^2]$ correctly, it is crucial to introduce the $\epsilon$ regulator in (\ref{eq:afterMellin}) and take it to zero at the very end. 

The source term at order $O(\v_a\v_c)$ that corresponds to the profiles (\ref{va}) is nontrivial. When  
combined with the results of the previous subsection \eqref{eq:gammasConstrained}, these lead to the constraints
\beq
\Lambda\[\widetilde{\mathcal{F}}'^{\mathbb{D}\mathbb{O}\mathbb{D}\mathbb{O}}_{jIiJ}(3) - \widetilde{\mathcal{F}}'^{\mathbb{D}\mathbb{O}\mathbb{D}\mathbb{O}}_{iIjJ}(1) +  \widetilde{\mathcal{H}}'^{\mathbb{O}\mathbb{D}\mathbb{D}\mathbb{O}}_{IjiJ}(3)-\widetilde{{H}}'^{\mathbb{O}\mathbb{D}\mathbb{D}\mathbb{O}}_{IijJ}(1)+\widetilde{\mathcal{H}}'^{\mathbb{O}\mathbb{D}\mathbb{D}\mathbb{O}}_{JijI}(3)-\widetilde{{H}}'^{\mathbb{O}\mathbb{D}\mathbb{D}\mathbb{O}}_{JjiI}(1)\]=2\delta_{ij}\delta_{IJ}\Delta + 2[M_{ij}]_{IJ}\,,
\eeq
and
\beq
\Lambda\[-\widetilde{\mathcal{F}}'^{\mathbb{D}\mathbb{O}\mathbb{D}\mathbb{O}}_{jIiJ}(2) + \widetilde{\mathcal{F}}'^{\mathbb{D}\mathbb{O}\mathbb{D}\mathbb{O}}_{iIjJ}(2) +  \widetilde{\mathcal{H}}'^{\mathbb{O}\mathbb{D}\mathbb{D}\mathbb{O}}_{IjiJ}(2)-\widetilde{{H}}'^{\mathbb{O}\mathbb{D}\mathbb{D}\mathbb{O}}_{IijJ}(-2)+\widetilde{\mathcal{H}}'^{\mathbb{O}\mathbb{D}\mathbb{D}\mathbb{O}}_{JijI}(2)-\widetilde{{H}}'^{\mathbb{O}\mathbb{D}\mathbb{D}\mathbb{O}}_{JjiI}(2)\]= [M_{ij}]_{IJ}\, .
\eeq
After plugging in the definitions of the Mellin transforms $\widetilde\cF$ (\ref{eqn-VLaplaceConvention}) and $\widetilde{\cal H}$ (\ref{Htilde}), we arrive at the non-homogeneous constraint (\ref{eqn-Constraint-InHomoFinal-1}) and (\ref{eqn-Constraint-InHomoFinal-2}) in the main text.

\subsection{Simplified Constraints for $\mathbb{O} = \mathbb{D}$}
\label{apd:KernelDDDDSimplified}

In the case where the operator $\mathbb O$ in (\ref{eqn-4ptDisDisOO}) is the displacement operator and $I=J=i=j$, we have a single four-point function to consider and we can use crossing symmetry to map all the integration regions to $t\in[0,1]$. The homogeneous and inhomogeneous constraints each reduce to a single independent nontrivial constraint. These take the form
\beq\la{eqn-Constraint-D1D1D1D1}
\int\limits_0^1 \dd t\,\mathcal{K}_{\mathtt{Homo}}(t)\mathcal{F}^{\DD \DD \DD \DD}_{1111}(t)=0\,,\qquad\Lambda\int\limits_0^1 \dd t\,\mathcal{K}_{\mathtt{Inhomo}}(t)\mathcal{F}^{\DD \DD \DD \DD}_{1111}(t)=4\,,
\eeq
where
\beq\la{DDDDkernels}
\mathcal{K}_{\mathtt{Homo}}(t)=1+t+t^2\,,\qquad\mathcal{K}_{\mathtt{Inhomo}}(t)=\left(t^2-{2t} - {2} \right) \log t +    \left({t^2} +{4}t + 1 \right) \log (1+t)\,.
\eeq

\section{OPE Sum Rules} \label{apd:OPESumRule}
The integrated constraints can be rephrased as OPE sum rules. In this appendix, we demonstrate how to convert between the two for the four-point function $\mathcal{F}^{\DD \DD \DD \DD}_{1111}(t)$. 
The homogeneous and inhomogeneous integral constraints \eqref{eqn-Constraint-D1D1D1D1} are of the form
\beq\la{DDDDconstraints}
\int\limits_0^1 \dd t\, \mathcal{K}(t) \mathcal{F}^{\DD \DD \DD \DD}_{1111}(t) = \frac{F}{\Lambda} \,,
\eeq
where $\mathcal{K}$ is one of the kernels in (\ref{DDDDkernels}) and $F$ is either zero or four. 

The OPE expansion of $\mathcal{F}^{\DD \DD \DD \DD}_{1111}(t)$ in the $t\to0$ channel takes the form
\beq\la{conformalblock}
\mathcal{F}^{\DD \DD \DD \DD}_{1111}(t) = \sum_{{\mathbb{O}}} {C}_{\DD \DD {\mathbb{O}}}^2\, \mathtt{Block1D}[\Delta_{{\mathbb{O}}}, t]\,,\qquad\text{with}\qquad\mathtt{Block1D}[\Delta, t] = \left(\frac{t}{1+t}\right)^\Delta {}_2F_1\left(\Delta, \Delta, 2\Delta, \frac{t}{1+t}\right) \,.
\eeq
This expansion converges for $t\in[0,1]$, \cite{Pappadopulo:2012jk}. Hence, after plugging it into the constraint (\ref{DDDDconstraints}) we can exchange the order of integration and summation. 
In this way, we arrive at
\beq
\sum_{{\mathbb{O}}}{C}_{\DD \DD {\mathbb{O}}}^2\,\widetilde{\mathtt{Block}}(\Delta_\mathbb{O})  = \Lambda F \,,\qquad\text{where}\qquad \widetilde{\mathtt{Block}}(\Delta) = \int\limits_0^1 \dd t\, \mathcal{K}(t)  \,\mathtt{Block1D}[\Delta, t]\,.
\end{equation}
For small $\Delta$ the integral may diverge. In such cases, it is evaluated by analytic continuation from larger $\Delta$. 

In the generic case where ${\mathbb O}\ne{\mathbb D}$ the four-point functions are integrated over the range $t\in[0,\infty]$. 
This integration domain can be split into the regions $t\in[0, 1]$ and $t\in[1, \infty]$.
In the first region, the $t\to0$ OPE expansion converges, and in the latter, the $t\to\infty$ OPE expansion converges. 
Using the convergent expansion in each region, we can 
exchange the order of summation and integration freely. In this way, we arrive at the sum rule of the form \eqref{eqn-SumRule}.

\section{A new prediction for $O(N)$ defect at order $\cO(\varepsilon^2)$} \label{apd:O(N)eps2}

In the main text, we have presented a prediction for $C_{\DD \DD \phi_1}^{(2)}$, see (\ref{eq:CDDphi1}). In this appendix, we use this result to obtain an integral constraint for the $\cO(\varepsilon^2)$ correction to the four-point function that does not involve any additional unfixed conformal data. 

We start by substituting $C_{\DD\DD\phi_1}^{(2)}$ from (\ref{eq:CDDphi1}) into the homogeneous constraints \eqref{eqn-Constraint-HomoFinal} and consider the result at order $O(\varepsilon^2)$. There are three sources of divergences, for the exchange of $\phi_1$, from the exchange of $\phi_1^2$, and from the exchange of $\phi_I^2$. The last two operators are degenerate at tree level and mix with each other at higher orders. We denote the two operators with a good anomalous dimension that result from this mixing by ${\mathbb S}_+$ and ${\mathbb S}_-$. Following the enhancement mechanism, we subtract from the four-point function the divergent contributions and separately add their finite coupling expressions before integration. We denote the subtracted four-point functions by $\mathcal{F}_{\mathtt{reg}}$. Since the singular behavior of the constraints depends on the form of the integration kernels, the subtraction differs between the first and second homogeneous constraints in \eqref{eqn-Constraint-HomoFinal}, and we treat them separately. For the first homogeneous constraint, we define the subtracted four-point function as
\beq
\label{eq:sub1}
\mathcal{F}^{\phi_1 \DD \DD \phi_1}_{\mathtt{reg},1} = \begin{dcases}
\mathcal{F}^{\phi_1 \DD \DD \phi_1} - C_{\DD \DD \phi_1}^2 \frac{1}{t^{\Delta_{\phi_1}}}  & \text{for}\ 0<t <1\\
\mathcal{F}^{\phi_1 \DD \DD \phi_1} - C_{\DD \DD \phi_1} C_{\phi_1 \phi_1 \phi_1} \frac{1}{t^{\Delta_{\phi_1}}}\Big(1-\frac{\Delta_{\phi_1}}{2 t} + \frac{(\Delta_{\phi_1} ^2+2 \Delta_{\phi_1} +1) \Delta_{\phi_1}}{4 (2 \Delta_{\phi_1} +1) t^2}\Big)  & \text{for}\ t>1\\
\qquad - \sum_{a = \pm } C_{\DD \DD \mathbb{S}_a} C_{\phi_1 \phi_1 \mathbb{S}_a} \frac{1}{t^{\Delta_{\mathbb{S}_a}}}\Big(1-\frac{\Delta_{\mathbb{S}_a}}{2 t}\Big) &
\end{dcases}\,,
\eeq
and 
\beq
\label{eq:sub2}
\mathcal{F}^{\DD \phi_1  \DD \phi_1}_{\mathtt{reg},1} = \begin{dcases}
\mathcal{F}^{\DD \phi_1  \DD \phi_1} & \text{for}\ 0<t <1\\
\mathcal{F}^{\DD \phi_1  \DD \phi_1} - C_{\DD \DD \phi_1}^2 t^{\Delta_{\phi_1}-4}  & \text{for}\ t>1\\
\end{dcases}\,,
\eeq
where the subtracted finite coupling terms are understood to be expanded at small $\varepsilon$ up to the same order as $\mathcal{F}$ before integration.

Substituting those definitions into the first homogeneous constraint and separately adding back the enhanced 
contributions from the $\cO(\varepsilon^3)$, we arrive at the constraint\footnote{Note that $C_{\phi_1 \phi_1 \mathbb{S}_a} = \cO(1)$ while $C_{\DD \DD \mathbb{S}_a} = \cO(\varepsilon)$.} 
\begin{align} \label{eqn-ONapd:homo1}
&\int\limits_0^\infty\dd t \Big[t^2(\mathcal{F}^{\phi_1 \DD \DD \phi_1}_{\mathtt{reg},1}+\mathcal{F}^{\DD \phi_1 \DD \phi_1}_{\mathtt{reg},1})+(1+t)^2\mathcal{F}^{\phi_1 \DD \DD \phi_1}_{\mathtt{reg},1} \Big]_{\cO(\varepsilon^2)} \\
=& \frac{2 \pi  C_{\DD \DD \phi_1}^{(2)}}{\sqrt{N+8}}-\frac{2 \pi  C_{\phi_1 \phi_1 \phi_1}^{(2)}}{3 \sqrt{N+8}}+\frac{7 \pi ^2}{3 (N+8)} + \sum_{a=\pm} \Big[ C_{\phi_1\phi_1\mathbb{S}_a}^{(0)} \left(2 \gamma_{\mathbb{S}_a}^{(1)} C_{\DD\DD \mathbb{S}_a}^{(1)}+\frac{9}{2} C_{\DD\DD \mathbb{S}_a}^{(2)}\right)+\frac{9}{2} C_{\DD\DD \mathbb{S}_a}^{(1)} C_{\phi_1 \phi_1 \mathbb{S}_a}^{(1)} \Big]\,.\nn
\end{align}

Similarly, for the second homogeneous constraint, we define the subtracted four-point function as
\beq
\mathcal{F}^{\phi_1 \DD \DD \phi_1}_{\mathtt{reg},2} = \begin{dcases}
\mathcal{F}^{\phi_1 \DD \DD \phi_1}, & \text{for}\ 0<t <1\\
\mathcal{F}^{\phi_1 \DD \DD \phi_1} - C_{\DD \DD \phi_1} C_{\phi_1 \phi_1 \phi_1} \frac{1}{t^{\Delta_{\phi_1}}}\Big(1-\frac{\Delta_{\phi_1}}{2 t} + \frac{(\Delta_{\phi_1} ^2+2 \Delta_{\phi_1} +1) \Delta_{\phi_1}}{4 (2 \Delta_{\phi_1} +1) t^2}\Big)  & \text{for}\ t>1\\
\qquad - \sum_{a = \pm } C_{\DD \DD \mathbb{S}_a} C_{\phi_1 \phi_1 \mathbb{S}_a} \frac{1}{t^{\Delta_{\mathbb{S}_a}}}\Big(1-\frac{\Delta_{\mathbb{S}_a}}{2 t}\Big)&
\end{dcases}\,,
\eeq
and $\mathcal{F}^{\DD \phi_1  \DD \phi_1}_{\mathtt{reg},2} =
\mathcal{F}^{\DD \phi_1  \DD \phi_1}$ with no subtractions. Substituting those definitions into the second homogeneous constraint and adding the enhanced contributions from the $\cO(\varepsilon^3)$, we arrive at
\begin{align} \label{eqn-ONapd:homo2}
&\int\limits_0^\infty\dd t \Big[2t\,t(1+t)\(\mathcal{F}^{\phi_1 \DD \DD \phi_1}_{\mathtt{reg},2}+\mathcal{F}^{\phi_1 \DD \DD \phi_1}_{\mathtt{reg},2}\)-2t\mathcal{F}^{\DD \phi_1 \DD \phi_1}_{\mathtt{reg},2}\Big]_{\cO(\varepsilon^2)} \\
&= \frac{\pi  C_{\DD \DD \phi_1}^{(2)}}{\sqrt{N+8}}\frac{\pi  C_{\phi_1 \phi_1 \phi_1}^{(2)}}{3 \sqrt{N+8}}-\frac{\pi ^2 \gamma_{\phi_1}^{(2)}}{N+8}+\frac{13 \pi ^2}{3 (N+8)}+ \sum_{a= \pm} \Big[ C_{\phi_1 \phi_1\mathbb{S}_a}^{(0)} (\gamma_{\mathbb{S}_a}^{(1)} C_{\DD\DD \mathbb{S}_a}^{(1)}+5 C_{\DD\DD\mathbb{S}_a}^{(2)})+5 C_{\DD\DD\mathbb{S}_a}^{(1)} C_{\phi_1\phi_1\mathbb{S}_a}^{(1)} \Big]\,.\nn
\end{align}

By taking a linear combination of \eqref{eqn-ONapd:homo1}  and \eqref{eqn-ONapd:homo2} so that unknown structure constant at this order, $C_{\phi_1 \phi_1 \phi_1}^{(2)}$, drops out we arrive at the constraint
\begin{align}\label{eqn-ONapd-finalRes}
&\int\limits_0^\infty\dd t\Big[t^2(\mathcal{F}^{\phi_1 \DD \DD \phi_1}_{\mathtt{reg},1}+\mathcal{F}^{\DD \phi_1 \DD \phi_1}_{\mathtt{reg},1})+(1+t)^2\mathcal{F}^{\phi_1 \DD \DD \phi_1}_{\mathtt{reg},1} + 2t\,t(1+t)\(\mathcal{F}^{\phi_1 \DD \DD \phi_1}_{\mathtt{reg},2}+\mathcal{F}^{\phi_1 \DD \DD \phi_1}_{\mathtt{reg},2}\)-2t\mathcal{F}^{\DD \phi_1 \DD \phi_1}_{\mathtt{reg},2}\Big]_{\cO(\varepsilon^2)}\\
&=\frac{  \pi^2  (2 \gamma^{(2)}_{\phi_1}-11)+{4 \pi  \sqrt{N+8}\,  C^{(2)}_{\DD\DD \phi_1}}}{N+8} + \sum_{a=\pm} \Big[4 \gamma_{\mathbb{S}_a}^{(1)} C_{\DD\DD\mathbb{S}_a}^{(1)} C_{\phi_1\phi_1\mathbb{S}_a}^{(0)}+\frac{29}{2} C_{\DD\DD\mathbb{S}_a}^{(2)} C_{\phi_1\phi_1\mathbb{S}_a}^{(0)}+\frac{29}{2} C_{\DD\DD\mathbb{S}_a}^{(1)} C_{\phi_1\phi_1\mathbb{S}_a}^{(1)} \Big]\,.\nn
\end{align}
Note that the OPE coefficient $C_{\phi_1\phi_1\mathbb{S}_a}$ up to $\cO(\varepsilon)$ can be found in \cite{Gimenez-Grau:2022czc}, while for the OPE coefficient $C_{\DD\DD\mathbb{S}_a}$ can be found in the expansion of the four-point function of $\mathcal{F}^{\phi_1 \DD \DD \phi_1}$. Therefore, the RHS of the equation consists of known CFT data and data that can be extracted from the four-point function at this order.\footnote{Note that $C_{\phi_1 \phi_1 \phi_1}^{(2)}$ cannot be extracted from $\mathcal{F}^{\phi_1 \DD \DD \phi_1}_{(2)}$ because it appears multiplied by $C_{\DD \DD \phi_1}$ which starts at order $O(\varepsilon)$.}

In more detail, the part subtracted in the $t \in [1, \infty)$ region takes the form
\begin{align}
t \in [1, \infty): \quad \mathcal{F}^{\phi_1 \DD \DD \phi_1}_{\mathtt{reg},1\text{ or }2}\Big|_{\cO(\varepsilon^2)} =&\ \mathcal{F}^{\phi_1 \DD \DD \phi_1}\Big|_{\cO(\varepsilon^2)}  - \varepsilon^2 \frac{\left(6 t^2-3 t+2\right) }{6 t^3} \left[C_{\phi_1\phi_1\phi_1}^{(1)} C_{\DD\DD\phi_1}^{(1)} \right] \\
& - \varepsilon^2 \frac{1}{t^2} \sum_{a = \pm} \left[ C_{\DD\DD\mathbb{S}_a}^{(2)} C_{\phi_1\phi_1\mathbb{S}_a}^{(0)}+C_{\DD\DD\mathbb{S}_a}^{(1)} C_{\phi_1\phi_1\mathbb{S}_a}^{(1)} -\log t\ \gamma_{\mathbb{S}_a}^{(1)}  C_{\DD\DD\mathbb{S}_a}^{(1)} C_{\phi_1\phi_1\mathbb{S}_a}^{(0)}\right]\nn\\
& - \varepsilon^2 \frac{1}{t^3} \sum_{a = \pm} \left[\frac{1}{2} \gamma_{\mathbb{S}_a}^{(1)} (2 \log t-1) C_{\DD\DD\mathbb{S}_a}^{(1)} C_{\phi_1\phi_1\mathbb{S}_a}^{(0)}  -C_{\DD\DD\mathbb{S}_a}^{(2)} C_{\phi_1\phi_1\mathbb{S}_a}^{(0)}-C_{\DD\DD\mathbb{S}_a}^{(1)} C_{\phi_1\phi_1\mathbb{S}_a}^{(1)}\right]\,.\nn
\end{align}
The first subtracted term corresponds to the $\phi_1$ exchange and depends only on the known OPE data from \eqref{ONdata}. After subtracting this term, the combinations appearing on the right-hand side of \eqref{eqn-ONapd-finalRes} can be extracted from the remaining $\mathcal{O}(1/t^2)$ terms: 
\begin{itemize}
    \item The coefficient of $\log t$ gives us the average $\sum_{a\pm} \gamma_{\mathbb{S}_a}^{(1)} C_{\DD\DD\mathbb{S}_a}^{(1)} C_{\phi_1\phi_1\mathbb{S}_a}^{(0)}$.
    \item The $\log t$-independent coefficient gives the average OPE coefficient $\sum_{a=\pm}\left[C_{\DD\DD\mathbb{S}_a}^{(2)} C_{\phi_1\phi_1\mathbb{S}_a}^{(0)}+C_{\DD\DD\mathbb{S}_a}^{(1)} C_{\phi_1\phi_1\mathbb{S}_a}^{(1)} \right]$.
\end{itemize}
Those data can also be extracted from the $\mathcal{O}(1/t^3)$ term.

\section{$\mathcal{N}=4$ Displacement OPE and Supercorrelator} \label{apd:N4OPE}

\subsection{Preliminary}

The 1/2 BPS Wilson loop in $\mathcal{N}=4$ SYM theory 
preserves an $\mathfrak{osp}(4^{*}|4)$ superconformal subgroup. Consequently, defect operators are labeled by their conformal dimension $\Delta$, 
their representation under the $\mathfrak{so}(3)$ transverse rotation group, and an $\mathfrak{sp}(4) \simeq \mathfrak{so}(5)$ $R$-symmetry Dynkin label $[a, b]$. In the following, we will label them by $[a,b]^{\Delta}_s$.

The displacement operator $\mathbb{D}_i$, with $i=1,2,3$ is the top component of the 1/2 BPS $\mathcal{D}_1$ supermultiplet, which consists of three states
\beq
\mathcal{D}_1: \quad [0,1]^{\Delta=1}_{s=0} \rightarrow [1,0]^{\Delta=3/2}_{s=1} \rightarrow [0,0]^{\Delta=2}_{s=2}\, .
\eeq
The OPE of two $\mathcal{D}_1$ supermultiplets contains the identity $\mathcal{I}$, the 1/2 BPS $\mathcal{D}_2$ supermultiplet, and an infinite set of unprotected \textit{scalar} long multiplets 
$\mathcal{L}^{\Delta}_{s=0, [0,0]}$
\beq \label{eqn-D1D1SuperOPE}
\mathcal{D}_1 \times \mathcal{D}_1 = \mathcal{I} + \mathcal{D}_2 + \sum_{\Delta} \mathcal{L}^{\Delta}_{s=0, [0,0]}\, .
\eeq
Here, the $\mathcal{D}_2$ multiplet consists of the states
\beq
\mathcal{D}_2: \quad[0,2]^{\Delta=2}_{s=0} \rightarrow[1,1]^{\Delta=\frac{5}{2}}_{s=1} \rightarrow[0,1]^{\Delta=3}_{s=2} \oplus[2,0]^{\Delta=3}_{s=0} \rightarrow[1,0]^{\Delta=\frac{7}{2}}_{s=1} \rightarrow[0,0]^{\Delta=4}_{s=0} \,.
\eeq
Details of those multiplets and their properties can be found in \cite{Agmon:2020pde,Ferrero:2023znz,Ferrero:2023gnu,Gunaydin:1990ag,Dorey:2018klg}. 

We turn to the structure of the four-point function $\llangle \mathcal{D}_1 \mathcal{D}_1 \mathcal{D}_1 \mathcal{D}_1 \rrangle$. Superconformal symmetry fixes this correlator up to an unknown function of the cross-ratio called \textit{the\ reduced\ correlator} $f(\chi)$
\beq \label{eqn-N4-SuperD1D1D1D1}
\frac{\llangle \mathcal{D}_1 \mathcal{D}_1 \mathcal{D}_1 \mathcal{D}_1 \rrangle}{\llangle \mathcal{D}_1 \mathcal{D}_1 \rrangle \llangle \mathcal{D}_1 \mathcal{D}_1 \rrangle} = \mathtt{F} \mathfrak{X} + \hat{\mathtt{D}} f(\chi)\,,
\eeq
where $\mathtt{F}$ is a constant, $\mathfrak{X}$ is the super cross-ratio, and $\hat{\mathtt{D}}$ is a differential operator. We refer the reader to \cite{Liendo:2018ukf} for their definition. In particular, in terms of the reduced correlator, crossing symmetry takes the form
\beq\la{crossingf}
\chi^2 f(\chi) = (1-\chi)^2 f(1-\chi)\,.
\eeq

To check the constraints, we consider the four-point function of the displacement operator. It takes the form\footnote{We thank Michelangelo Preti for kindly sharing a $\mathtt{Mathematica}$ code for the superspace computations.}
\beq \label{eqn-apd-N4-DDDDgen}
\llangle \mathbb{D}_i(x_1)  \mathbb{D}_j(x_2) \mathbb{D}_k(x_3) \mathbb{D}_l(x_4) \rrangle = \frac{1}{x_{12}^4 x_{34}^4} \Big[ \delta_{ij}\delta_{kl} \mathcal{F}_S(\chi) + \( \delta_{ik}\delta_{jl}-\delta_{il}\delta_{jk} \) \mathcal{F}_A(\chi) +\( \delta_{ik}\delta_{jl}+\delta_{il}\delta_{jk} - \frac{2}{3}\delta_{ij}\delta_{kl} \) \mathcal{F}_T(\chi) \Big]\,,
\eeq
where the cross ration $\chi$ is related to our cross ratio $t$ as  $\chi = \frac{t}{1+t}$, and 
\beq
\begin{aligned}
\label{eqn-DDDDcorFromReduced}
\mathcal{F}_S(\chi) = & -\frac{1}{108} (\chi -1)^2 ((\chi -3) \chi +3) \chi ^4 f^{(5)}(\chi )-\frac{1}{108} (\chi -1) (\chi  (\chi  (16 \chi -33)+3)+18) \chi ^3 f^{(4)}(\chi ) \\
& -\frac{1}{54} (\chi  (\chi  (\chi  (37 \chi -54)+9)-36)+45) \chi ^2 f^{(3)}(\chi )+\frac{1}{3} \left(-3 \chi ^4+\chi ^3-6 \chi +9\right) \chi  f''(\chi )\\
&+\left(-\frac{\chi ^4}{3}+4 \chi -7\right) f'(\chi )+\left(\frac{8}{\chi }-4\right) f(\chi )+\frac{\mathtt{F} \chi ^4}{3}\, , \\
\mathcal{F}_A(\chi) = & -\frac{1}{72} (\chi -2) (\chi -1)^2 \chi ^5 f^{(5)}(\chi )-\frac{1}{36} (\chi -1) (\chi  (8 \chi -13)+3) \chi ^4 f^{(4)}(\chi ) \\
& -\frac{1}{36} (\chi  (\chi  (37 \chi -53)+9)+6) \chi ^3 f^{(3)}(\chi )+\frac{1}{4} \left(-6 \chi ^5+3 \chi ^4+2 \chi ^2\right) f''(\chi )\\
&-\frac{1}{2} \left(\chi ^3+2\right) \chi  f'(\chi )+f(\chi )+\frac{\mathtt{F} \chi ^4}{2}\, , \\
\mathcal{F}_T(\chi) = & -\frac{1}{72} (\chi -1)^2 \chi ^6 f^{(5)}(\chi )-\frac{1}{18} (\chi -1) (4 \chi -3) \chi ^5 f^{(4)}(\chi ) \\
& -\frac{1}{36} (\chi  (37 \chi -51)+15) \chi ^4 f^{(3)}(\chi )+\frac{1}{4} (5-6 \chi ) \chi ^4 f''(\chi )-\frac{1}{2} \chi ^4 f'(\chi )+\frac{\mathtt{F} \chi ^4}{2}\, .
\end{aligned}
\eeq

The $\mathcal{D}_1 \mathcal{D}_1$ OPE shown in \eqref{eqn-D1D1SuperOPE} implies that the reduced correlator can be expanded as:
\beq \label{eqn-D1D1SuperOPEBlock}
f(\chi) = f_{\One} + C_{\mathcal{D}_1\mathcal{D}_1\mathcal{D}_2}^2 f_{\mathcal{D}_2}+ \sum_{\mathbb{O}} C_{\mathcal{D}_1\mathcal{D}_1\mathbb{O}}^2 f_{\mathbb{O}}\, ,
\eeq
where the terms $f_{\mathbb{I}}$, $f_{\mathcal{D}2}$, and $f_{\mathbb{O}}$ are referred to as \textit{reduced blocks}, serving as direct analogs of the standard conformal blocks. As shown in \cite{Liendo:2018ukf,Ferrero:2023znz, Ferrero:2023gnu}, supersymmetry fixes their form to be
\beq \label{eqn-D1D1SuperOPEBlockExpression}
\begin{aligned} 
    f_{\One} & = \chi, \\
    f_{\mathcal{D}_2} & = \chi -\chi  \, {}_2F_1(1,2;4;\chi ) = O(\chi^2)\, , \\
    f_{\mathbb{O}} & = 
    \frac{\chi ^{\Delta_{\mathbb{O}}+1} \, _2F_1(\Delta_{\mathbb{O}}+1,\Delta_{\mathbb{O}}+2;2 \Delta_{\mathbb{O}}+4;\chi )}{1-\Delta_{\mathbb{O}}} = O(\chi^{\Delta_{\mathbb{O}}+1})\, .
\end{aligned}
\eeq

At a generic value of the YM coupling, there are no $\mathfrak{sp}(4)$ singlets operators with integer dimension $\Delta<4$. This fact is sufficient for our constraints to hold, see section \ref{sec:singint}. We test them below using the CFT data equated above.

\subsection{Simplifing the Homogeneous Constraint} \label{apd-N4-homoGen}

Here, we will show that the homogeneous constraints fix the leading OPE behavior of the reduced correlator. 
After subtracting $\cF_\text{GFF}$, we plug the form of the correlator \eqref{eqn-DDDDcorFromReduced} into the homogeneous constraints (\ref{eqn-Constraint-HomoFinal}) and perform multiple integrations by parts to remove the derivatives from the reduced correlator $f(\chi)$. After this, we find that the integrations over the reduced correlator reduce to a set of boundary terms. We are left with the integration of a topological term, $\mathtt{F} \mathfrak{X}$ in (\ref{eqn-N4-SuperD1D1D1D1}), and the homogeneous constraint reduces to 
\beq\la{homoBPS}
\int\limits_0^1 d \chi\ \mathtt{F} \( \chi^\ell \delta_{ij}\delta_{IJ}  +  \chi^{2-\ell} \delta_{iI}\delta_{jJ} +  (\chi -1)^{\ell} \chi ^{2-\ell} \delta_{iJ}\delta_{jI} \) =\mathtt{[boundary\ terms]}\,,
\eeq
where $\ell=0,1,2$ corresponds to the three homogeneous constraints in \eqref{eqn-Constraint-HomoFinal}. We have verified that (\ref{homoBPS}) is satisfied, 
provided that the reduced correlator has the following asymptotics behavior close to $\chi=0$ and $\chi=1$ 
\beq\la{eqn-ReducedF-Bdr}
f(\chi) = - \mathtt{F} \chi^2/2 + \cO(\chi^3)\,, \qquad f(\chi) = \mathtt{F}/2 + \cO\left((1-\chi)^2\right)\, .
\eeq
These two behaviors are related by crossing. They are the result of the exchange of the ${\cal D}_2$ multiplet, see (\ref{eqn-D1D1SuperOPEBlockExpression}). The other operators in the OPE (\ref{eqn-D1D1SuperOPE}) are scalar long multiplets. They give contributions of order $\cO(\chi^{\Delta_{\mathbb O}+1})$. These are subleading due to the unitarity bound, $\Delta_{\mathbb{O}} > 1$. This 
asymptotic behavior fixes the value of the protected OPE coefficient, defined in \eqref{eqn-D1D1SuperOPEBlock}, as  $C_{\mathcal{D}_1\mathcal{D}_1\mathcal{D}_2}^2 = \mathtt{F}$. It agrees with the finite-coupling asymptotics in \cite{Liendo:2018ukf}, where this structure constant is fixed by supersymmetric localization. 

\subsection{Simplifying the Inhomogeneous Constraint} \label{apd-N4-InhomoGen}

As before, we subtract from (\ref{eqn-N4-SuperD1D1D1D1}) $\cF_\text{GFF}$ and plug the result into the inhomogeneous constraints (\ref{eqn-Constraint-InHomoFinal-1}) and (\ref{eqn-Constraint-InHomoFinal-2}). We then perform integrations by parts to remove the derivatives from the reduced correlator. We first consider the constraint (\ref{eqn-Constraint-InHomoFinal-1}) and contract it with $\delta_{Ii}\delta_{Jj}$. Since there may be potential divergences at the integration boundaries, we introduce a cut-off $\epsilon_1$ near $\chi=1$ and $\epsilon_0$ near $\chi=0$. After several integrations by parts, the equation reduces to
\begin{equation}\la{firstinhom}
\int\limits_{\epsilon_0}^{1-\epsilon_1} \dd \chi \[ 2f(\chi)\left( \frac{1}{\chi} - \frac{1}{\chi^3} + \frac{1}{3 \chi^2} + \frac{1}{3(1-\chi)} \right) 
\] + \mathtt{[boundary\ terms]}={4\over\Lambda}\,.
\end{equation}

Using (\ref{eqn-ReducedF-Bdr}) from the homogeneous constraint and the explicit form of the $\mathtt{F} \mathfrak{X}$ term, the boundary terms take the form
\begin{equation} \label{eqn-N4-reducedBdrTerm}
\mathtt{[boundary\ terms]} =-\frac{4\mathtt{F}}{3}
+ \frac{\mathtt{F}}{3} \log (\epsilon_1) + \mathtt{F} \log (\epsilon_0) + \cO(\epsilon_0, \epsilon_1) \,.
\end{equation}
Here, the last two cutoff-dependent terms can be written as 
\beq \label{eqn-DDDD-divBdrTerm}
\frac{\mathtt{F}}{3} \log (\epsilon_1) + \mathtt{F} \log (\epsilon_0) = - \frac{\mathtt{F}}{3} \int\limits_{\epsilon_0}^{1-\epsilon_1} \frac{d \chi}{1-\chi} - \mathtt{F} \int\limits_{\epsilon_0}^{1-\epsilon_1} \frac{d \chi}{\chi} + \cO(\epsilon_0, \epsilon_1)\, .
\eeq
Adding them back to the bulk integration, we observe that it becomes \textit{finite} due to the boundary asymptotics given in \eqref{eqn-ReducedF-Bdr}.  As a result, the equation reduces to
\beq\la{eqsimp1}
 \int\limits_0^1 \dd \chi \[ 2f(\chi)\left( \frac{1}{\chi} - \frac{1}{\chi^3} + \frac{1}{3 \chi^2} + \frac{1}{3(1-\chi)} \right) 
- \frac{\mathtt{F}}{3} \frac{1}{1-\chi} - \frac{\mathtt{F}}{\chi} 
-\frac{4\mathtt{F}}{3}\] ={\Lambda\over4}\,.
\eeq

To further simplify the result, we notice that the crossing (\ref{crossingf}) implies that
\beq
\int\limits_0^1 d \chi\, f(\chi) = \int\limits_0^1 d \chi\, \frac{f(\chi)}{\chi}\,, \qquad  \int\limits_0^1 d \chi\, \frac{f(\chi)}{\chi^2}  = 0\, .
\eeq
Thus we can rewrite (\ref{eqsimp1}) as
\beq \label{eqn-N4ConstraintFiniteCoupling}
\frac{4}{3} \int\limits_0^1 d \chi\, \[f(\chi)\left( 1 - \frac{2}{\chi^3} \right) -\(1+\frac{1}{\chi}\)\mathtt{F} \]= {4\over\Lambda}\,,
\eeq
where we have simultaneously performed the change of variable $\chi \rightarrow 1-\chi$ for the terms with $1-\chi$ in the denominator. This form of the equation is finite due to (\ref{eqn-ReducedF-Bdr}). This constraint (\ref{eqn-N4ConstraintFiniteCoupling}) is in agreement with a relation that was found and tested in \cite{Cavaglia:2022yvv,Drukker:2022pxk}. 

The analysis of the other inhomogeneous equations follows a similar procedure. We find that they reduce to multiples of \eqref{eqn-N4ConstraintFiniteCoupling}. The only exception is the $\delta_{iI}\delta_{jJ}$ term in the first inhomogeneous constraint, which, in addition to involving multiples of \eqref{eqn-N4ConstraintFiniteCoupling}, also depends on two subleading Taylor coefficients of the reduced correlator near $\chi = 0$. Given \eqref{eqn-N4ConstraintFiniteCoupling}, these new contributions must cancel among themselves. This leads to the condition
\beq\la{extraconstraint}
6 a_3  - 4 a_4 + \mathtt{F} = 0\,, \quad \text{where} \quad f(\chi) = -\frac{\mathtt{F}}{2} \chi^2 + a_3 \chi^3 + a_4 \chi^4 + \cO(\chi^5)\, .
\eeq

At large coupling the dimension of long operators is large. Therefore, they do not contribute to the small $\chi$ OPE expansion up to order $\cO(\chi^4)$. Since $f(\chi)$ is expected to be analytic in the 't Hooft coupling, they do not contribute to (\ref{extraconstraint}) at finite coupling. As a result, the coefficients $a_3$ and $a_4$ receive contributions only from the protected multiplet $\mathcal{D}_2$. 
We find that the expansion of the reduced block, $f_{\mathcal{D}2}$ in \eqref{eqn-D1D1SuperOPEBlockExpression}, indeed satisfies this relation.

In conclusion, for the 1/2 BPS line in ${\cal N}=4$ SYM theory, the constraints agree with the available CFT data. 

\section{$1/2$ BPS defects in $AdS_3 \times S^3 \times T^4$}
\label{app:ads3}

Type IIB string theory in $AdS_3 \times S^3 \times T^4$ is dual to a $2$ dimensional CFT. The geometric background $AdS_3 \times S^3 \times T^4$ can support both Ramond-Ramond (RR) and Neveu-Schwarz–Neveu-Schwarz (NS-NS) three-form fluxes. For our purposes, it suffices to know that the mixed fluxes are parameterized by a real angle $\lambda \in [0,\pi/2]$ that extrapolates between the pure RR ($\lambda=0$) and the pure NS-NS ($\lambda = \pi/2$) cases. This parameter represents a true moduli of the dual CFT. Additionally, the theory depends on the string coupling $g_s$ which is related to an integer parameter, $N$, in the CFT\footnote{To be more precise, the string theory is dual to a grand canonical ensemble of CFTs with different $N$, \cite{Eberhardt:2021jvj}.} and the ratio between the string scale and the AdS radius that is related to a sort of 't Hooft coupling in the dual CFT, $g= R^2_\text{AdS}/(2 \pi l_s^2)$.

In \cite{Bliard:2024bcz}, the authors have studied a classical string configuration in this background that is holographically dual to a $\frac{1}{2}$-BPS line defect in the 't Hooft limit where the CFT is strongly coupled. 
We will test the constraints to leading nontrivial order at strong coupling, $g \rightarrow \infty$. 

At the leading order in $1/g$, the four-point function of the displacement operator is given by $\cF_\text{GFF}$ (\ref{eq:GFFpiece}), which we subtract. 
The next order, $\cO(1/g)$, has been computed in two different ways. One is based on tree-level diagrams in the effective theory on the classical $AdS_2$ string. The other is done using analytical bootstrap methods. 
The analytical bootstrap result depends on two undetermined parameters, $b_1$ and $b_2$, and takes the following form
\begin{equation} \label{eqn-AdS3-Fboostrap}
\begin{split}
    g \mathcal{F}_{1111} = &-\frac{2 (t (t+1) (t (t (t (2 t+9)+16)+14)+6)+2)  (b_1-2 b_2) \log (t)}{3 (t+1)^5}\\
    & -\frac{4b_1 \left(t^2+t+1\right)\left(t (t+1) \left(t \left(3 (t+2) t^2+t-2\right)+9\right)+3\right)}{9 (t+1)^4t^4}\\
    & +\frac{2b_2 \left(t^2+t+1\right) \left(t (t+1) \left(t \left(12 (t+2) t^2+t-11\right)+36\right)+12\right)}{9 (t+1)^4t^4}\\
    &+\frac{2 \left(2 t^6+t^5+t+2\right) (b_1-2 b_2) \log (t+1)}{3 t^5}\,,
\end{split}
\end{equation}
where $g$ enters the bootstrap through
\beq
\Lambda=\frac{6 g}{\pi} +\cO\left(g^0\right)\,.
\eeq

The effective string result, on the other hand, depends on only one parameter, $\lambda$.
By comparison of the two results, \cite{Bliard:2024bcz} determine,
\beq
\label{eq:b1b2lambda}
b_1+4b_2+6/\pi=0\,.
\eeq
In the following, we find that our integrated constraints can also be used to fix this relation without using the result of the holographic computation. Note that without additional input, any bootstrap computation cannot fix a moduli of a CFT such as $\lambda$. Its relation to these parameters is found to be $b_1=-2 \sin ^2(\lambda )/\pi$. 

A new feature that appears in this case is that the OPE of two displacement operators includes the displacement operator itself. That is because the NS-NS flux breaks parity and thus reflection in the transverse direction is not a symmetry of the straight defect. This fact is reflected in the small-$t$ expansion of (\ref{eqn-AdS3-Fboostrap})
\beq \label{eqn-AdS3-1111-smallLimit}
g\mathcal{F}_{1111}(t) = -\frac{b_1}{3} \left( t^{-2} - t^{-1} + \cO(t^0) \right)\, ,
\eeq
where the first two terms match with the expansion of the conformal block of a dimension two operator. 

We have repeated the procedure described in the main text for the exchange of an operator of dimension $\Delta$ and send $\Delta\to2$ at the end. We find that the homophonous constraints are satisfied while the inhomogeneous constraints reduce to the condition 
(\ref{eq:b1b2lambda}). Hence, the constraints are compatible with both the string theory and the bootstrap computations. 

\section{Conformal Block in One Dimension}

The generic scalar four-point function can be expanded as \cite{Dolan:2000ut, Dolan:2003hv}
\beq \label{eqn-apd-4ptCanonical} 
\llangle \mathbb{O}_{\Delta_1}(x_1)\mathbb{O}_{\Delta_2}(x_2)\mathbb{O}_{\Delta_3}(x_3)\mathbb{O}_{\Delta_4}(x_4) \rrangle =  \left( \frac{x_{24}}{x_{14}}\right)^{\Delta_2 -\Delta_1}\left( \frac{x_{14}}{x_{13}}\right)^{\Delta_3 -\Delta_4} \frac{\widehat{\mathcal{F}}(\chi)}{x_{12}^{\Delta_1 + \Delta_2}x_{34}^{\Delta_3 + \Delta_4}}\,,
\eeq
where
\beq
\chi = \frac{x_{12}x_{34}}{x_{13}x_{24}} = \frac{t}{1+t}\,,
\eeq
is the cross-ratio. It is convenient to work with this cross-ratio in this section, as it simplifies the differential equation of the conformal block.

The function $\widehat{\mathcal{F}}(\chi)$ can be expanded in conformal blocks. In the $S$-channel, the expansion is given by:
\beq
\widehat{\mathcal{F}}(\chi) = \sum_{\widetilde{\mathbb{O}}} C_{\mathbb{O}_{\Delta_1} \mathbb{O}_{\Delta_2} \widetilde{\mathbb{O}}}C_{\widetilde{\mathbb{O}}\mathbb{O}_{\Delta_3} \mathbb{O}_{\Delta_4}} \mathtt{Block1D}_{\Delta_{\widetilde{\mathbb{O}}}}(\chi)\,,
\eeq
where the one-dimensional conformal block can be expressed in terms of the hypergeometric ${}_2 F_1$ function as \cite{Jackiw:2012ur,Ghosh:2023wjn}
\beq
\mathtt{Block1D}_{\Delta}(\chi) = \chi^{\Delta}\, {}_2 F_1(\Delta-\Delta_1+\Delta_2, \Delta+\Delta_3-\Delta_4; 2\Delta; \chi)\, .
\eeq
For our purposes, it is sufficient to consider the conformal block for a scalar intermediate operator.

Since our normalization of the four-point function differs from that used in \eqref{eqn-apd-4ptCanonical}, we must account for this change. Specifically, we have:
\beq
\mathtt{Block1D}_{\Delta}^{\mathbb{O}\DD\DD\mathbb{O}}(\chi) = \mathtt{Block1D}_{\Delta}^{\DD\mathbb{O}\DD\mathbb{O}}(\chi)  = (1-\chi )^4 \chi ^{\Delta -\Delta_{\mathbb{O}}-2} \, _2F_1(\Delta -\Delta_{\mathbb{O}}+2,\Delta -\Delta_{\mathbb{O}}+2;2 \Delta ;\chi )\, . 
\eeq
These are equal because we have rescaled $\mathcal{F}^{\mathbb{D} \mathbb{O} \mathbb{D} \mathbb{O}}$ as defined in \eqref{funnyFs}.

In the main text, we will also consider the expansion in the $T$-channel where we take $x_1 \rightarrow x_4$ and $x_2 \rightarrow x_3$, 
\beq
\widehat{\mathcal{F}}(\chi) = \sum_{\widetilde{\mathbb{O}}} C_{\mathbb{O}_{\Delta_2} \mathbb{O}_{\Delta_3} \widetilde{\mathbb{O}}}C_{\widetilde{\mathbb{O}}\mathbb{O}_{\Delta_4} \mathbb{O}_{\Delta_1}} \overline{\mathtt{Block1D}}_{\Delta_{\widetilde{\mathbb{O}}}}(\chi)\,,
\eeq
the conformal blocks $\overline{\mathtt{Block1D}}$ can be obtained from $\mathtt{Block1D}$ by crossing, which takes
\beq
x_i \rightarrow x_{(i+1)\ \mathtt{mod}\ 4}, \quad \chi \rightarrow 1-\chi, \quad \Delta_i \rightarrow \Delta_{(i+1)\ \mathtt{mod}\ 4}\, .
\eeq

Taking into account the change in the prefactor from \eqref{eqn-apd-4ptCanonical}, the crossed conformal block is given by:
\beq
\overline{\mathtt{Block1D}}_{\Delta}(\chi) = \chi ^{\Delta_3+\Delta_4} (1-\chi )^{-\Delta_2-\Delta_3+\Delta } \, _2F_1(\Delta -\Delta_2+\Delta_3,\Delta -\Delta_1+\Delta_4;2 \Delta ;1-\chi )\, .
\eeq
For the four-point function of interest, the corresponding conformal blocks in the crossed channel, after accounting for the change in prefactor from \eqref{eqn-4ptDisDisOO} and \eqref{funnyFs}, read:
\beq \label{eqn-1DblockCross}
\begin{aligned}
\overline{\mathtt{Block1D}}_{\Delta}^{\mathbb{O}\DD\DD\mathbb{O}}(\chi) & = (1-\chi )^{\Delta } \, _2F_1(\Delta ,\Delta ;2 \Delta ;1-\chi )\, , \\
\overline{\mathtt{Block1D}}_{\Delta}^{\DD\mathbb{O}\DD\mathbb{O}}(\chi) & = (1-\chi )^{\Delta -\Delta_{\mathbb{O}}+2} \, _2F_1(\Delta -\Delta_{\mathbb{O}}+2,\Delta +\Delta_{\mathbb{O}}-2;2 \Delta ;1-\chi )\, .
\end{aligned}
\eeq

\end{document}